\documentclass[aps,prb,twocolumn,superscriptaddress,floatfix,nobibnotes]{revtex4-1}

\pdfoutput=1

\usepackage{ifpdf}

\usepackage[american]{babel}   %%% an english thesis usually has a german synopsis %ngerman
\usepackage[T1]{fontenc}  

\usepackage{exscale} %%% for proportional sum and integral signs

\usepackage{braket} %Bras und Kets 
\usepackage{multirow} 
\usepackage{ellipsis} 
\usepackage{xspace} 
\usepackage{cancel} 

\usepackage{booktabs} %%% scientific publication style tables
\usepackage{empheq}   %%% for boxed equations and alike
\usepackage{xcolor}   %%% for colored text and formulae
\usepackage{amsmath}  %%% math symbols needed
\usepackage{amsfonts} %%% math fonts needed
\usepackage{amssymb}  %%% maths symbols needed 
\usepackage{amsthm}   %%% theorem environments
\usepackage{dsfont}   %%% for blackboard numbers and characters 
\usepackage{upgreek}

\usepackage[pdftex]{graphicx}       %%% CUSTOMIZATION: include all pictures
\usepackage{float}      %%% define custom float environments

\usepackage[
colorlinks=true
]{hyperref} %%% CUSTOMIZATION use this package for the final compilation run for online publication, includes hyperlinks for sections/eqs/etc.

\usepackage{subfigure}     %%% nice captioning and referencing of subfigures
\newcommand{\I}{i} 
\newcommand{\e}{\mathrm e} 
\newcommand{\id}{\mathds{1}} 
\newcommand{\dd}{\mathrm{d}}             %%% an operator d for total derivatives and integrals

\DeclareMathOperator{\re}{Re}
\DeclareMathOperator{\im}{Im}

\DeclareMathOperator{\cth}{coth}
\DeclareMathOperator{\sign}{\mathrm{sign}} 
\DeclareMathOperator{\tr}{Tr} 
\newcommand{\nel}{n_{\mathrm{el}}}				%%% electron concentration
\newcommand{\myint}{\displaystyle\int\limits}	%%% integral with boundaries
\newcommand{\myintl}{\displaystyle\int}			%%% integral with boundaries
\newcommand{\mysum}{\displaystyle\sum\limits}	%%% sum with indices
\newcommand{\action}{\mathcal{S}}				%%% action
\newcommand{\C}{\mathcal{C}}					%%% dimensionless parameter
\newcommand{\Z}{\mathcal{Z}}					%%% partition function
\newcommand{\D}{\mathcal{D}}					%%% functional integral
\newcommand{\Psiq}{\overline{\Psi}}				%%% conjugate Grassmann field
\newcommand{\vecphi}{{\boldsymbol \phi}}			%%% real vector field
\newcommand{\vecafm}{\mathbf{Q}}					%%% AFM vector
\newcommand{\vecr}{\mathbf{r}}						%%% space coordinate
\newcommand{\vecR}{\mathbf{R}}						%%% space coordinate
\newcommand{\vecq}{\mathbf{q}}						%%% q-vector
\newcommand{\veck}{\mathbf{k}}						%%% k-vector
\newcommand{\pauli}{\boldsymbol \sigma}				%%% Pauli matrix
\newcommand{\cutoffk}{k_{\Lambda}}				%%% momentum cutoff
\newcommand{\cutoffe}{E_F}						%%% highest energy
\newcommand{\Nhs}{\mathcal{N}}					%%% number of lin. ind. AFM vectors 
\newcommand{\eff}{\mathrm{eff}}					%%% effective
\newcommand{\ian}{\mathrm{int}}					%%% interaction
\newcommand{\ds}{\displaystyle}					%%% displaystyle
\newcommand{\crosstl}{T^{\ast}}					%%% lower crossover 
\newcommand{\crosstu}{\tilde{T}}				%%% upper crossover
\newcommand{\zz}{z}								%%% rescaled momentum
\newcommand{\xz}{\zeta}							%%% rescaled frequency
\newcommand{\tf}{f}								%%% function x \coth x/2
\newcommand{\hh}{\mathrm{hs}}					%%% hot spots
\newcommand{\cp}{\mathrm{c}}					%%% composite modes
\newcommand{\nuf}{\nu_F}						%%% DOS at Fermi level
\newcommand{\g}{\sqrt{u}}						%%% spin-fermion coupling constant
\newcommand{\gsq}{u}							%%% squared spin-fermion coupling constant
\newcommand{\gqu}{u^2}							%%% quartic spin-fermion coupling constant
\newcommand{\lphi}{L_{\varphi}}					%%% phase relaxation length
\newcommand{\drude}{\sigma_0}					%%% Drude conductivity
\newcommand{\Ceff}{\C_{\mathrm{eff}}}			%%%effective composite interaction
\newcommand{\phim}{\Phi}						%%% product of \phi and Pauli matrix
\newcommand{\cond}{g_0}							%%% dimensionless conductance

\newcommand{\ie}{\mbox{i.\,e.,}~\nolinebreak[4]} %%% 'that is': "For reasons not fully understood there is only a minor PSI contribution to the variable fluorescence emission of chloroplasts (Dau, 1994), i.e. the PSI fluorescence appears to be independent from the state of its reaction centre (Butler, 1978)."
\newcommand{\cf}{\mbox{cf.}~\nolinebreak[4]} %%% 'confer', meaning "compare with": "These results were similar to those obtained using different techniques (cf. Wilson, 1999 and Ansmann, 1992)."

\begin{document}

\title{Interference of quantum critical excitations and soft diffusive modes in a disordered antiferromagnetic metal}

\author{Philipp~S.~Wei\ss }
\affiliation{Institut f\"ur Theorie der Kondensierten
Materie, Karlsruher Institut f\"ur Technologie, D-76131 Karlsruhe, Germany}
\altaffiliation[Present address: ]{Institute for Theoretical Physics, University of Cologne, D-50937, Cologne, Germany}
\author{Boris~N.~Narozhny }
\affiliation{Institut f\"ur Theorie der Kondensierten
Materie, Karlsruher Institut f\"ur Technologie, D-76131 Karlsruhe, Germany}
\affiliation{National Research Nuclear University MEPhI (Moscow Engineering
Physics Institute), 115409 Moscow, Russia}
\author{J\"{o}rg~Schmalian}
\affiliation{Institut f\"ur Theorie der Kondensierten
Materie, Karlsruher Institut f\"ur Technologie, D-76131 Karlsruhe, Germany}
\affiliation{Institut f\"ur Festk\"orperphysik, Karlsruher Institut
f\"ur Technologie, D-76021 Karlsruhe, Germany}
\author{Peter~W\"{o}lfle}
\affiliation{Institut f\"ur Theorie der Kondensierten
Materie, Karlsruher Institut f\"ur Technologie, D-76131 Karlsruhe, Germany}
\affiliation{Institut f\"ur Nanotechnologie, Karlsruher Institut f\"ur Technologie,
D-76021 Karlsruhe, Germany}

\date{\today}

\begin{abstract}
We study the temperature-dependent quantum correction to conductivity due to the interplay of spin density fluctuations and weak disorder for a two-dimensional metal near an antiferromagnetic (AFM) quantum critical point.
AFM spin density fluctuations carry large momenta around the ordering vector $\vecafm$ and, at lowest order of the spin-fermion coupling, only scatter electrons between ``hot spots'' of the Fermi surface which are connected by $\vecafm$.
Earlier, it was seen that the quantum interference between AFM spin density fluctuations and soft diffusive modes of the disordered metal is suppressed, a consequence of the large-momentum scattering.
The suppression of this interference results in a non-singular temperature dependence of the corresponding interaction correction to conductivity.
However, at higher order of the spin-fermion coupling, electrons on the entire Fermi surface can be scattered successively by two spin density fluctuations and, in total, suffer a small momentum transfer.
This higher-order process can be described by composite modes which carry small momenta.
We show that the interference between formally subleading composite modes and diffusive modes generates singular interaction corrections which ultimately dominate over the non-singular first-order correction at low temperatures.
We derive an effective low-energy theory from the spin-fermion model which includes the above-mentioned higher-order process implicitly and show that for weak spin-fermion coupling the small-momentum transfer is mediated by a composite propagator.
Employing the conventional diagrammatic approach to impurity scattering, we find the correction $\delta \sigma \propto +\ln^2 T$ for temperatures above an exponentially small crossover scale.
\end{abstract}

\maketitle

%\section{Introduction}

Quantum interference plays a crucial role in the electronic transport
of disordered metals. At low temperatures, the conductivity is largely
dominated by elastic scattering of electrons off static disorder.
A classical description of impurity scattering leads to the well-known
Drude conductivity ${\drude = e^2 \nel \tau/m}$, where $\tau$ is the
transport mean-free time, and $e$, $\nel$, and $m$ are the charge,
density, and mass of the electrons, respectively. Taking into account
interference processes, one finds corrections to $\drude$, which are
typically small, but exhibit strong temperature dependence~\cite{ALTb85}.
Already at the one-particle level, interference of
time-reversed trajectories leads to the weak localization
correction~\cite{Gorkov1979,Altshuler1980,Abrahams1979}.  In the
presence of electron-electron interaction, coherent scattering off
Friedel oscillations results in the Altshuler-Aronov
corrections~\cite{Finkelstein1983,Finkelstein1984,ALTb85,ZNA01}.  In
two dimensions (2D) and at low temperatures (where electron motion is
diffusive), both types of corrections are logarithmic,
${\delta\sigma\propto\ln T}$.

Tuning the system to the proximity of a second-order quantum phase
transition, one typically finds that the physics is determined
by critical fluctuations~\cite{SAC11}.
If at the same time weak disorder is present, then
such critical fluctuations may interfere with the diffusive modes
leading to enhanced temperature dependence of the quantum corrections
to transport coefficients.
%%%%%%%%% CHANGES
Examples of such phenomena were discussed
in Ref.~\onlinecite{Kim2003} in the context of a metamagnetic quantum critical point (QCP)
and
in Ref.~\onlinecite{PAU05} in the context of the ferromagnetic QCP.
In the latter case, 
%%%%%%%%%%%%
the paramagnetic phase possesses a
critical region close to the quantum phase transition which is dominated by the critical
fluctuations of the spin density. This system can be described in
terms of the spin-fermion model~\cite{ABA03}, which treats the spin
fluctuations and low-energy electrons independently. Moreover, the
spin fluctuations mediate the effective electron-electron interaction,
which for small momenta is more singular than the usual Coulomb
potential. Consequently, coherent scattering off Friedel oscillations
is enhanced: the Altshuler-Aronov correction evaluated with this
effective interaction exhibits squared logarithmic behavior,
${\delta\sigma\propto\ln^2 T}$ (in the 2D case). Similar behavior was
seen by Ludwig~\textit{et~al.}~\cite{Ludwig2008} in the effect of gauge-field
interaction on fermion transport in two dimensions.

\begin{figure}[t]
	\centering
	\includegraphics[width=8.6cm]{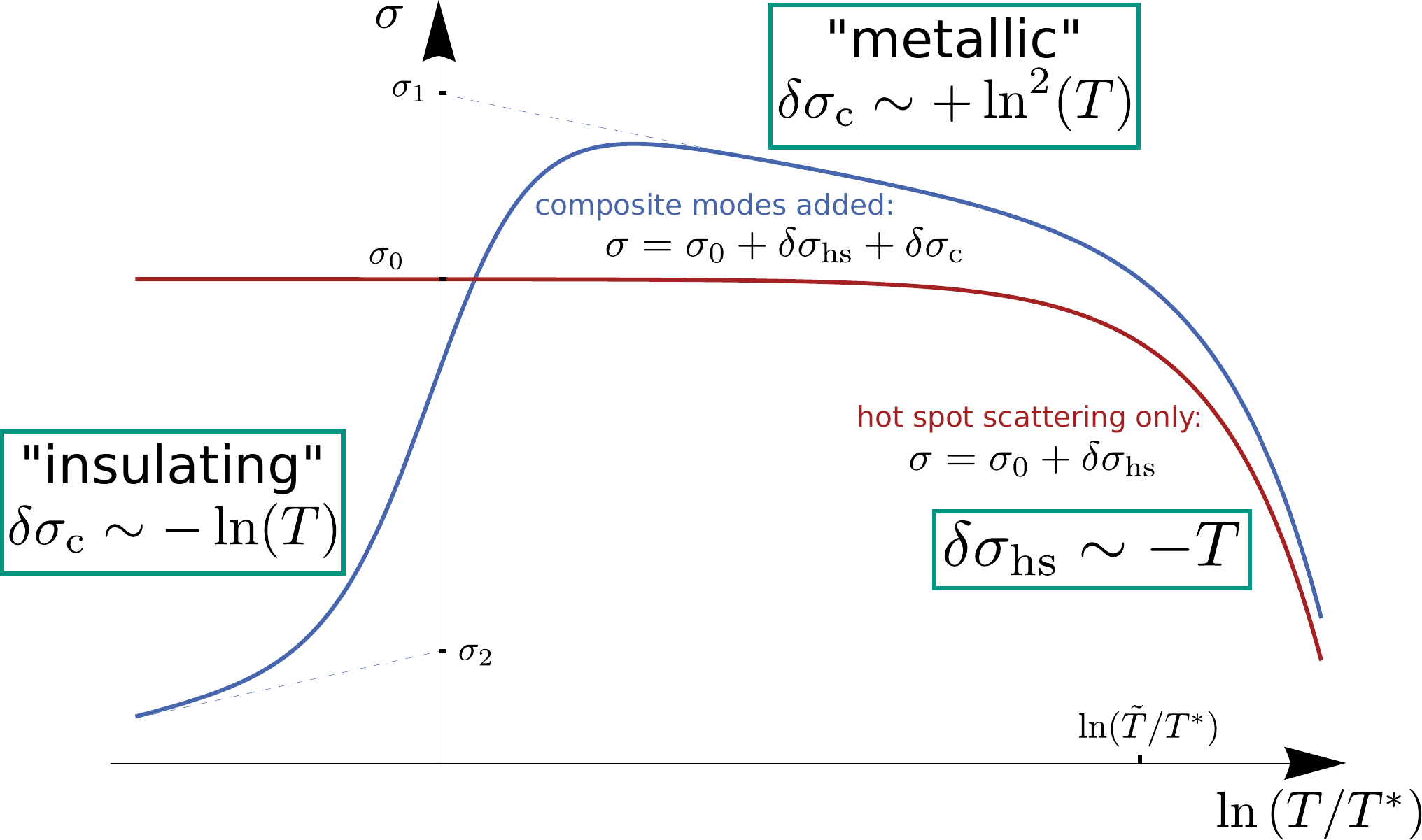}
        \caption{Temperature dependence of the
          interaction correction to conductivity in a 2D disordered
          metal near an AFM QCP. Red curve: Drude
          conductivity with the interaction correction due to hot-spot
          scattering~\cite{SYZ12},
          ${\sigma=\drude+\delta\sigma_{\hh}}$. Blue curve: modified
          temperature dependence once the correction due to the
          composite modes is taken into account,
          ${\sigma=\drude+\delta\sigma_{\hh}+\delta\sigma_{\cp}}$. The
          dashed extrapolated curves intersect the vertical axis at
          ${\sigma_1=\drude+e^2/(8\pi^2\lambda\hbar)}$ and
          ${\sigma_2=\drude-e^2/(2\pi^2\lambda\hbar)}$. The crossover
          temperatures $\crosstl$ and $\crosstu$ are defined in the
          main text.}
	\label{fig:sigma-plot}
\end{figure}

The situation at an antiferromagnetic (AFM) QCP is
quite different. Here, critical spin density fluctuations carry large
momenta of the order of the AFM ordering wave vector $\vecafm$. As a
result, via leading-order processes electrons can be scattered by the AFM fluctuations only
between few special points on the Fermi surface, the so-called ``hot
spots.'' One might then conclude, that the discrepancy between typical
momentum scales of the AFM fluctuations and the diffusive modes will
lead to nonsingular temperature dependence of the
Altshuler-Aronov--type interaction correction~\cite{SYZ12}. In this
paper, we show that such conclusion would be premature.
The reason is the emergence of a composite collective mode~\cite{HAR11,ABRa14}
that, while of subleading order in the dimensionless coupling constant,
is singular at small momenta.
Interference between these
composite modes and the diffusive modes leads to a correction to
the conductivity that exhibits a singular temperature dependence (see
Fig.~\ref{fig:sigma-plot}), and thus is more relevant at low
temperatures than the leading-order hot spot scattering~\cite{SYZ12}.

Theoretical investigations of the resistivity of disordered
AFM metals go back to 1977, when Ueda~\cite{UED77}
considered spin density fluctuations around $\vecafm$ in his
calculation of the resistivity. Treating electron scattering off
potential disorder and the spin density fluctuations on equal footing
within the Boltzmann equation approach, he found an expression for the
correction to the
resistivity of a nearly antiferromagnetic, disordered metal in three
dimensions, ${\delta\rho\propto T^{3/2}}$. Twenty years later, the problem of
the quasiclassical resistivity of disordered AFM metals was revisited
by Rosch~\cite{ROS99}, who focused on the dichotomy of the ``hot''
and ``cold'' manifolds of the Fermi surface
and analyzed three- and quasi-two-dimensional systems.
While in clean systems
the less resistive cold sections of the Fermi surface have been argued~\cite{HLU95}
to ``short-circuit'' the contribution of
quasiparticles at hot spots, Rosch demonstrated that impurity
scattering broadens the hot spots and recovered the ${T^{3/2}}$
temperature dependence (in contrast to the $T^2$ behavior found in the
clean case~\cite{HLU95}).

Quantum corrections in the disordered AFM metals close to the QCP were
recently considered by Syzranov and Schmalian in
Ref.~\onlinecite{SYZ12} within the spin-fermion model. It was shown
that the interference processes involving the AFM spin density
fluctuations and the diffusive modes are suppressed due to the large-momentum
transfer in the hot-spot scattering. Technically, at each
spin-fermion vertex the arguments of the electronic Green's functions
are shifted by ${\sim E_F}$ relatively to each other. Therefore, each
additional impurity line results in the small factor
${1/(E_F\tau)\ll1}$. The corresponding interaction correction to
resistivity due to the hot-spot scattering was found to be ${\delta
\rho \propto T^{d/2} }$ in $d$ dimensions, in full agreement with the
Boltzmann theory. Thus, diffusive modes appeared to be irrelevant for
transport at the AFM QCP.

However, for higher-order processes the argument of
Ref.~\onlinecite{SYZ12} breaks down. While the contribution of such
processes to thermodynamic quantities is less singular compared to the
leading-order results, transport phenomena are affected by the
higher-order processes in a qualitatively different way. Following
Hartnoll~\textit{et~al.}~\cite{HAR11}, we consider the effect of multiple
scattering of electrons off the spin density fluctuations. In the
simplest process, a fermion is scattered by two spin density
fluctuations successively via an intermediate off-shell state, such
that the momentum difference between the initial and final electronic
states (near the Fermi surface) can be arbitrarily small. Such process
can be described in terms of scattering off a composite mode that
combines both spin fluctuations. Recently, such composite modes were
shown to renormalize the quasiparticle mass in a strong coupling 
theory of critical spin density fluctuations~\cite{ABRa14}.

The above composite modes mediate an effective electron-electron
interaction with small-momentum transfer. This raises the question of
whether the interference processes involving this effective
interaction and the diffusive modes would give rise to singular
corrections to the conductivity. Here we report a positive answer to this
question (see Fig.~\ref{fig:sigma-plot}). Our argument consists of two
major steps. First, we demonstrate the emergence of the composite
modes in perturbation theory and find the effective electron-electron
interaction. Second, we follow the standard calculation of the
interaction corrections~\cite{ALTb85,ZNA01} in order to find the
effect of the critical composite modes on electronic transport in 2D
disordered AFM metals close to the QCP.

Our findings are illustrated in Fig.~\ref{fig:sigma-plot} where we
plot the conductivity near an AFM critical point as a function of
temperature. The dominant (in amplitude) contribution to the
conductivity stems from the Drude expression $\drude$. The red curve
shows the temperature dependence of the conductivity found in
Ref.~\onlinecite{SYZ12}, ${\sigma=\drude+\delta\sigma_{\hh}}$, where
the quantum correction $\delta\sigma_{\hh}$ arises due to scattering
of electrons between the hot spots induced by the critical spin
density fluctuations. The resulting conductivity is decreasing
linearly with temperature, ${\delta\sigma_{\hh}\propto -T}$ (which on the
logarithmic scale of Fig.~\ref{fig:sigma-plot} shows as the
exponential drop). This behavior should be contrasted with our main
result, i.e., the temperature dependence of the correction
$\delta\sigma_{\cp}$ resulting from the interference processes
involving the composite modes. This correction exhibits two distinct
regimes. In a wide intermediate temperature range, we find an
antilocalizing correction with a stronger than usual temperature
dependence, ${\delta\sigma_{\cp}\propto+\ln^2T}$. For the lowest
temperatures, this behavior is replaced by the expected localizing
behavior, ${\delta\sigma_{\cp}\propto -\ln T}$. The crossover between
the two regimes occurs at ${T^\ast=\cutoffe \e^{-1 / \lambda} }$,
which is exponentially small for small effective coupling constant
$\lambda$. The overall temperature dependence of the conductivity
comprising both types of corrections is shown by the blue curve in
Fig.~\ref{fig:sigma-plot}. We estimate that below a certain
temperature scale,
${\tilde{T} }$
%%%% CHANGES
the $\ln^2T$-correction exceeds the
%%%%
leading-order result ${\delta\sigma_{\hh}}$.

%%%%CHANGES
%As in Ref.~\onlinecite{SYZ12} we consider disorder consisting of non-magnetic impurities,
%which create a random potential, \ie propagation of the electrons. 
We describe the disorder in the metal by a random potential acting on the electrons only.
In contrast, if magnetic impurities are present, both spin fluctuations and
magnetic moments of the electrons couple directly to the disorder.
Similar to a single Kondo impurity in a normal metal, one expects
a significant $T$ dependence of the conductivity.
In this paper, we restrict ourselves to the simpler case of non-magnetic impurities,
as it was done in Ref.~\onlinecite{SYZ12}.
%Departing from the same model, we are focused on comparing our findings
%to the previously obtained results.
%%%%%%

The remainder of the paper is organized as follows. In
Sec.~\ref{afm-sfm}, we introduce the spin-fermion model for the AFM
metal close to the QCP.  The composite modes and the resulting
effective interaction are described in Sec.~\ref{comp-modes}. In
Sec.~\ref{sec:imp-model}, we introduce the quenched disorder. The
subsequent Sec.~\ref{int-cor} details the calculation of the
interaction correction. The result is compared to the previously
known expressions 
%%%%% CHANGE
and applied to experimental resistivity data
%%%%%%%%
in Sec.~\ref{disc}. 
The closing arguments are
presented in Sec.~\ref{concl}.

\section{AFM spin-fermion model}
\label{afm-sfm}

	In principle, the spin-fermion model is the result of a renormalization procedure where fermionic high-energy degrees of freedom are integrated out from a microscopic lattice model in order to obtain an effective low-energy theory.
	We do not attempt a rigorous derivation and, instead, place qualitative arguments in order to motivate the model.
	
\begin{figure}[t]
	\centering
	\includegraphics[width=6.0cm]{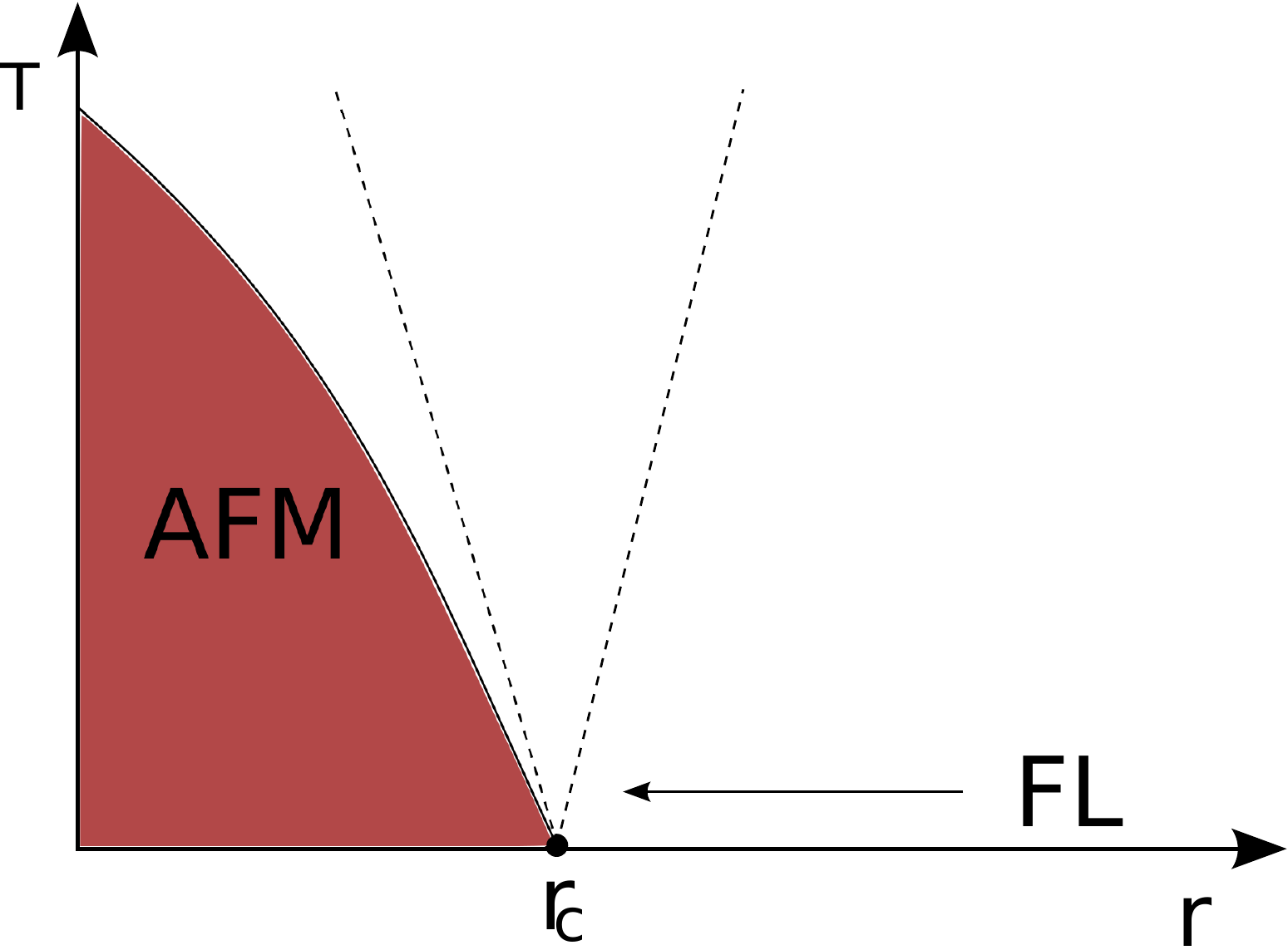}
	\caption{Typical phase diagram of an AFM metal. $T$ denotes the temperature and $r$ represents a non-thermal tuning parameter.
	For ${r< r_c}$, the electron spins are antiferromagnetically ordered with respect to two sublattices.
	For large values of ${r> r_c}$, the electron spins are disordered and the electrons are described by the Fermi liquid (FL) theory.
	If the system is tuned close to the transition (following the arrow in the phase diagram) spin density fluctuations develop, ultimately becoming soft modes.
	Within the cone above the QCP (indicated by the dashed lines), the critical fluctuations are thermally populated.
	}
	\label{fig:phase-diagram-thesis}
	\end{figure}

	An AFM metal exhibits an antiferromagnetically ordered phase below a transition temperature $T_N$.
	In the ordered phase, the spins of the electrons, responsible for the itinerant magnetism, point in opposite directions on two atomic sublattices.
	In the antiferromagnetically ordered phase, the magnetic order parameter is non-zero and spatially modulated according to ${\braket{\mathbf{M}_j} \sim \e^{\I \vecafm\cdot \vecR_j}}$
	where $\vecR_j$ is a lattice vector.
	The periodicity of the order parameter is given by the AFM ordering wave vector $\vecafm$.
	In reciprocal space, this ordering maps to the magnetic Brillouin zone which is spanned 
	by $\vecafm$.
	Above $T_N$, in the paramagnetic phase, the electron spins are disordered.
	By applying pressure, a magnetic field or by changing the chemical composition the transition temperature can be tuned to zero.
	If this transition is of second order, a QCP separates the antiferromagnetic and the paramagnetic phase.
	A typical phase diagram of an AFM metal is depicted in Fig.~\ref{fig:phase-diagram-thesis}.
	Aside from the temperature $T$, $r$ represents the non-thermal tuning parameter.
	By changing the tuning parameter the metal can undergo a quantum phase transition from the paramagnetic phase to the AFM phase at ${T=0}$.
	If the paramagnetic electron system approaches the QCP at ${r=r_c}$, spin density fluctuations pronounce the transition to the antiferromagnetically ordered phase and, ultimately, become soft modes.
	Typically, at ${T>0}$ the critical fluctuations persist in thermally excited states in a cone above the QCP in the phase diagram, indicated by the dashed lines in Fig.~\ref{fig:phase-diagram-thesis}.
	
	Our treatment of the AFM quantum phase transition relies on the following basic assumptions, similar to Ref.~\onlinecite{ABA03}:
	First, we presume the existence of the AFM QCP of the two-dimensional metal at ${T=0}$. 
	The AFM ordering is described by an AFM ordering wave vector $\vecafm$ with a magnitude comparable to the Fermi momentum, ${|\vecafm| \approx k_F}$.
	Even though the vector $\vecafm$ and the Fermi momentum are independent quantities, their magnitudes scale with the lattice spacing.
	Therefore, it seems to be reasonable to assume that they are of the same order.
	Second, the high-energy degrees of freedom, which are responsible for the antiferromagnetism,
	do not affect the low-lying fermionic excitations, \ie the quasiparticle picture is still valid close to the AFM instability.
	Besides, critical spin density fluctuations emerge near the AFM second-order quantum phase transition.
	
	The spin-fermion model describes the metal close to the AFM quantum phase transition in terms of fermionic quasiparticles and critical spin density fluctuations~\cite{ABA03},
	\begin{equation} \label{eq:action_spin-fermion-model-mom}	
	\begin{array}{rcl}
	\action 		& = & 	\dfrac{1}{\beta}\mysum_{k} \Psiq_k G^{-1}_{0,k} \Psi_k 
							+\dfrac{1}{\beta}\mysum_{q} \left| \vecphi_{q} \right|^2 \chi^{-1}_{0,q} \\ \\
					%
%					&	&	+\dfrac{\g}{\beta^2 L}\mysum_{k_1 k_2} \Psiq_{k_1} \left[ \pauli \cdot \vecphi_{k_1-k_2} \right] \Psi_{k_2} \Theta\left(q^\ast-|\veck_1-\veck_2-\vecafm|\right)
					&	&	+\dfrac{\g}{\beta^2 L}\mysum_{k q} \Psiq_{k+q} \left[ \pauli \cdot \vecphi_{q} \right] \Psi_{k} \,,
					%

%					&	&	+\dfrac{\g}{\beta^2 L}\mysum_{q k} \Psiq_{k+q} \left[ \pauli \cdot \vecphi_{k_1-k_2} \right] \Psi_{k_2} \vartheta\left(q^\ast-|\veck_1-\veck_2-\vecafm|\right)\,.
	\end{array} 
	\end{equation}
	where ${\Psi_k = (\Psi_{k\uparrow}, \Psi_{k\downarrow}  )}$ is a spinor of Grassmann fields, $\vecphi_q$ is a three-component order-parameter field, and ${\pauli~=~(\sigma^x,\sigma^y,\sigma^z)}$ are the Pauli matrices.
	The summation over imaginary frequencies is implicit in our notation, \ie ${k = (\I \epsilon_n, \veck)}$ and ${q = (\I\omega_m, \vecq)}$ for fermions and bosons, respectively.
	While the fermionic quasiparticles are described by the free-fermionic Green's function $G_{0,k}$, 
	the spin susceptibility $\chi_{q}$ plays the role of the propagator of the spin density fluctuations.
	The critical excitations are subject to a spin-fermion interaction with coupling constant $\g$.
	As spin density fluctuations are precursors of the AFM order,
	the peak of the spin susceptibility lies at the AFM ordering wave vector $\vecafm$ which is typically a large vector, comparable to the Fermi momentum.
	As a consequence, spin density fluctuations carry large momenta.
	More specifically, including the effect of low-energy Landau damping we use the following form of the renormalized spin susceptibility:
	\begin{equation} \label{eq:spin-sus-form}	
	\begin{array}{rcl}
	\chi_{q}^{-1} & = & \xi^{-2} + \left( \vecq \mp \vecafm \right)^2 + \gamma |\omega_m| + \dfrac{\omega_m^2}{c^2}\,.
	%\chi_{q}^{-1} & = & \xi^{-2} + |\vecq\,|^2 + \gamma |\omega_m| + \dfrac{\omega_m^2}{c^2}
	\end{array} 
	\end{equation}
	Without AFM long-range order, the spin susceptibilities are equal for each component of the order-parameter field.
	$\vecq$ is the momentum and ${\vecq \mp \vecafm}$ measures the small deviation from the large vector ${\pm \vecafm}$.
	Formally, we introduce a cutoff $q^\ast$ for the deviations and declare ${q^\ast / |\vecafm| \ll 1}$ a small parameter of our theory.
	$\xi$~is the correlation length which diverges at the critical point.
	Right at the critical point the gap in the excitation spectrum vanishes and the spin susceptibility is singular for small deviations from ${\vecq = \pm \vecafm}$.
	We neglect the mass term close to the critical point and set ${\xi^{-2} =0}$ from now on.
	$\gamma$ is a phenomenological damping constant and $c$ is a velocity.

\begin{figure}[t]
	\centering
	\includegraphics[width=5.0cm]{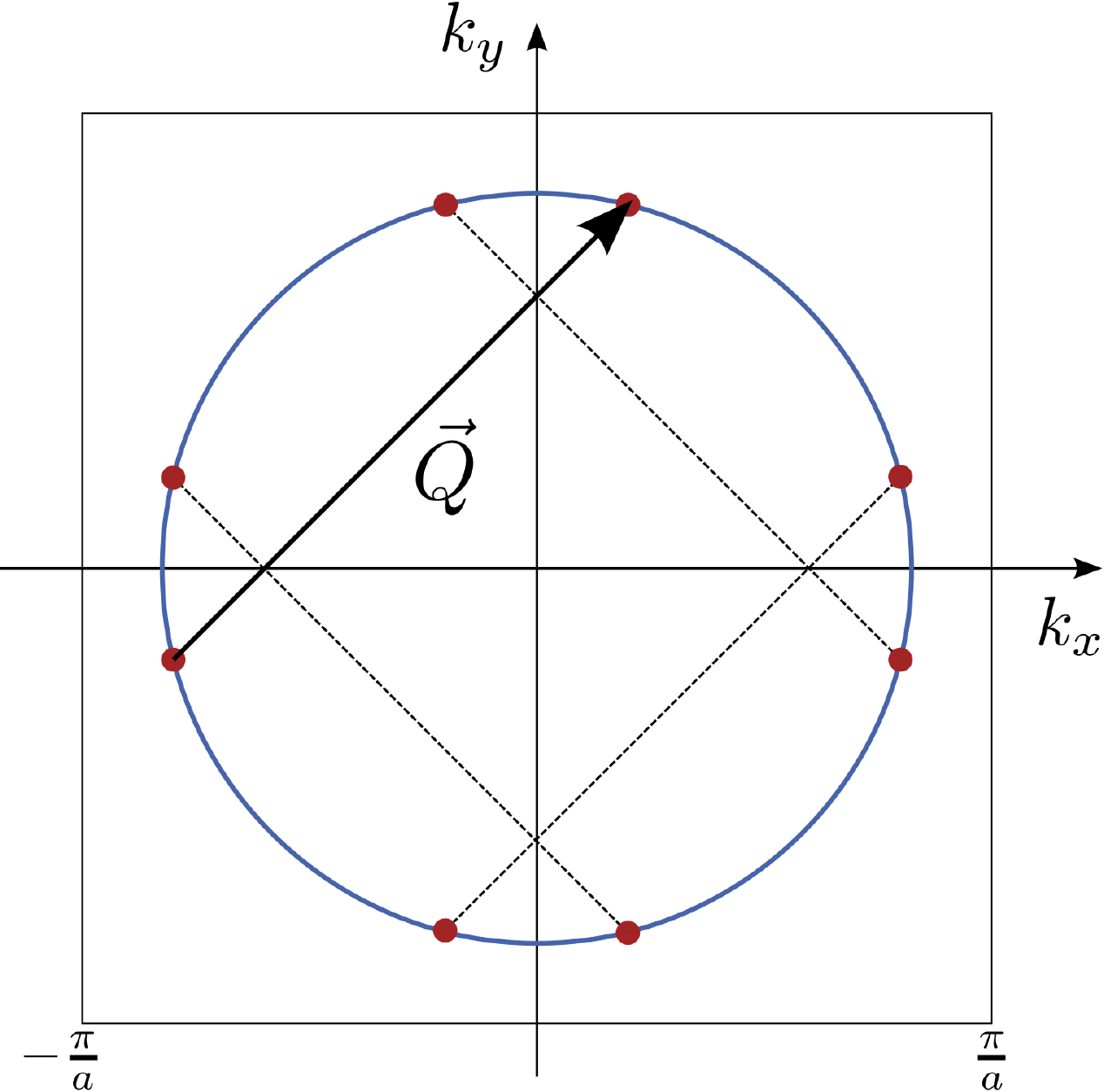}
	\caption{In reciprocal space, the magnetic unit cell of an antiferromagnetically ordered square lattice maps to the magnetic Brillouin zone which is spanned by the AFM ordering wave vector ${\vecafm=( \pi /a, \pi /a)}$.
	Carrying momenta around $\vecafm$, spin density fluctuations scatter electrons only between 8 hot spots.
	For simplicity, we draw a generic circular Fermi line.
	We emphasize that the calculations of this paper are not restricted to a specific form of $\vecafm$. 
	}
	\label{fig:scattering-hot-spots-thesis}
	\end{figure}

	At low temperatures, the fermionic modes are bound to a thin shell around the Fermi surface and we cut off the summation over $\veck$ at a distance $\cutoffk$ from the Fermi momentum $k_F$ where ${\cutoffk / k_F \ll 1}$.
	The restriction of fermionic momenta to the Fermi surface together with the restriction of bosonic momenta to the respective AFM wave vector $\vecafm$ limits the phase space of the spin-fermion coupling.
	The initial and final fermionic momenta $\veck$ and $\veck+\vecq$ are restricted by the cutoff $\cutoffk$ to the Fermi surface and the momentum transferred by the spin density fluctuations $\vecq$ is restricted by the cutoff $q^\ast$ to the vicinity of $\vecafm$.
	As a consequence of momentum conservation, at the lowest order $\gsq$, spin density fluctuations can scatter fermions only between small patches of the Fermi surface which are connected by $\vecafm$.
	These patches are called ``hot spots'' since they are coupled to spin density fluctuations prominently while the remaining Fermi surface stays ``cold'' in this sense.
	As an example, Fig.~\ref{fig:scattering-hot-spots-thesis} illustrates the hot spots on the Fermi line for a generic spherical Fermi surface and the AFM ordering vector ${\vecafm = (\pi /a, \pi /a)}$ of a square lattice with a magnetic unit cell of size $a$.
	However, the results of this paper are not restricted to a specific form of $\vecafm$.
	
	The spin density fluctuations are damped due to the interaction with fermions at the hot spots.
	In order to account for the damping we introduce the phenomenological damping constant $\gamma$.
	In fact, $\gamma$ is related to the spin-fermion coupling constant.
	If the damping is due to particle-hole excitations at the hot spots, the damping constant is determined by the polarization operator ${\Pi(\vecafm,\I\omega_m) - \Pi(\vecafm,0) \sim -\gsq |\omega_m|}$.
	Yet, we do not specify the damping mechanism and treat $\gamma$ as an independent input parameter.

\section{Composite modes from the AFM spin-fermion model}
\label{comp-modes}

\begin{figure}[t]
	\centering
	\includegraphics[width=8.6cm]{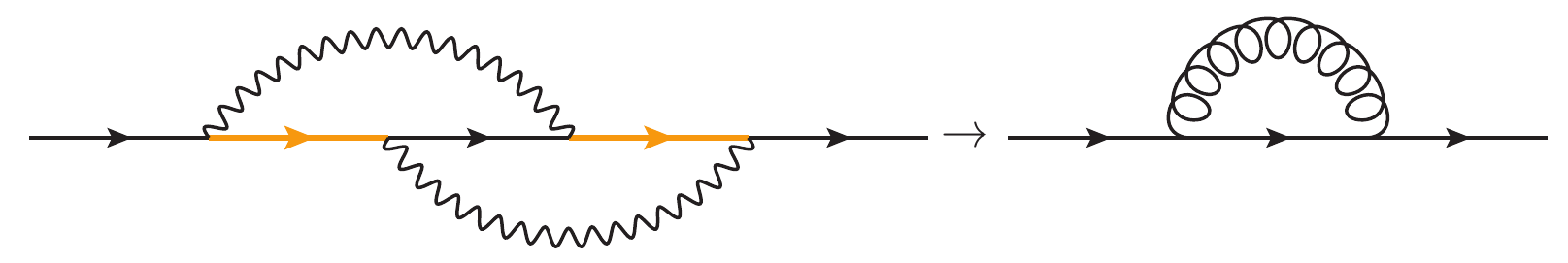}
	\caption{Self-energy correction to the fermionic Green's function at order $\gqu$.
	The orange lines denote Green's function of high-energy fermions shifted by $E_F$ from the Fermi surface, the thin solid lines denote low-energy Green's functions of on-shell fermions.
	The wavy lines represent propagators of spin density fluctuations which are joint in the definition of a composite propagator, represented by the curly line.
	}
	\label{fig:idea-composite-propagator}
	\end{figure}

In higher-order processes AFM spin density fluctuations can transfer small momenta although a single fluctuation carries the large momentum $\vecafm$.
	A fermion at an arbitrary point of the Fermi surface can be scattered by a spin density fluctuation to an intermediate off-shell state and, then, be scattered back to the Fermi surface.
	In total, the fermion suffers only a small momentum transfer.
	As an example, we consider a diagram in the perturbative expansion of the fermionic Green's function beyond the order of $\gsq$.
	At order $\gqu$, the self-energy corrections are represented by diagrams with two wavy lines of spin density fluctuations.
	One of these diagrams is depicted in Fig.~\ref{fig:idea-composite-propagator}.
	As the fermionic Green's functions are shifted by a large vector $\vecafm$ at each spin-fermion vertex, the fermionic Green's functions between the first and the second and between the third and the fourth vertex necessarily describe high-energy fermions for generic external momenta $\veck$.
	If the first spin density fluctuation carries $\vecafm+\vecq$, the second fluctuations carries $-\vecafm+\vecq_1$ and returns the fermion back to the Fermi surface.
	$\vecq$ and $\vecq_1$ denote the small deviations from $\vecafm$ from now on.
	In total, the double scattering transfers the small momentum $\vecq + \vecq_1$.
	In the following, we argue that both spin density fluctuations can be combined to a composite mode which effectively carries small momenta.
	
	The interaction correction to the Drude conductivity corresponding to such higher-order processes is parametrically smaller in the spin-fermion coupling $\sim\gqu$ as the interaction correction due to the hot spot scattering $\sim\gsq$.
	However, in a disordered metal the small-momentum-scattering process may be boosted by diffusive modes.
	We ask the question whether the interplay of composite modes and diffusive modes leads to qualitatively different and relevant corrections to the Drude conductivity.
	
\subsection{Effective low-energy theory for weak spin-fermion coupling}	

	\subsubsection{Scale separation and integrating out high-energy modes} \label{sec:scale-separation}
		In this section, we derive an effective low-energy theory from the spin-fermion model which includes scattering of fermions via intermediate off-shell states implicitly.
		To this end, we extend the fermion sector from the momentum shell near the Fermi surface up to the bandwidth $W$ of the underlying lattice model.
		The bandwidth $W$ is assumed to be sufficiently large, such that all fermionic intermediate states which can be reached by $\vecafm$ from the Fermi surface lie within the band.
		Formally, we neglect the finite size of the band and take $W\rightarrow \infty$. 
		The fermion sector is divided into low-energy modes and high-energy modes with respect to the momentum cutoff $\cutoffk$ of the original spin-fermion model,
		\begin{equation} \label{eq:definition_low_high_modes} 
		\begin{array}{rcl}
		\Psi^{<}_{k} 				& \equiv & \Theta_{\cutoffk	-|k_{\perp}|} \, \Psi_{k}\,, \\ \\
		\Psi^{>}_{k} 				& \equiv & \Theta_{|k_{\perp}|-\cutoffk}  \, \Psi_{k}\,, \\ \\
		\end{array} 
		\end{equation}
		where $k_{\perp}$ measures the distance from the Fermi surface.
		$\Theta_x$ is the usual step function.
		In order to avoid any complication of nested Fermi surfaces we assume a generic spherical Fermi surface.
		The extended action of the spin-fermion model, in terms of high-energy modes and low-energy modes, is conveniently written in matrix form, similar to a two-level system:
		\begin{equation} \label{eq:critical_action_<_>_matrix} 
		\begin{array}{rcl}
		\action & = & \action_{\phi} 
					+ \dfrac{1}{\beta^2}\mysum_{k_1 k_2} \mysum_{a,b = <,> }
					\overline{\Psi}^{a}_{k_1}
					\left[G^{a b}_{k_1 k_2} \right]^{-1}
					\Psi^{b}_{k_2} \,. \\ \\

%		%
%		\action		& = & 	
%		%
%		\action_{\phi}
%		%
%		+ \dfrac{1}{\beta^2}\mysum_{k_1 k_2}
%		%
%		\begin{pmatrix} \overline{\Psi}^{<}_{k_1} & \overline{\Psi}^{>}_{k_1} \end{pmatrix}
%		%
%		\begin{pmatrix}
%		\left[G^{< <}_{k_1 k_2} \right]^{-1} & \left[G^{< >}_{k_1 k_2} \right]^{-1} \\ \left[G^{> <}_{k_1 k_2} \right]^{-1} & \left[G^{> >}_{k_1 k_2} \right]^{-1}
%		\end{pmatrix}
%		%
%		\begin{pmatrix} \Psi^{<}_{k_2} \\ \Psi^{>}_{k_2} \end{pmatrix} 
%		%
		\end{array} 
		\end{equation}
		The matrix elements read as
		\begin{equation} \label{eq:critical_action_<_>_matrix_elements} 
		\begin{array}{rcl}
		\left[G^{< <}_{ k_1 k_2} \right]^{-1} & = & 
		\beta \left[ G^{<}_{0,k_1}\right]^{-1} \updelta_{k_1 k_2}
			+ \g \,\phim_{k_1 -k_2} \,,\\ \\
		\left[G^{> >}_{ k_1 k_2} \right]^{-1} 	& = & 
		\beta \left[ G^{>}_{0,k_1}\right]^{-1} \updelta_{k_1 k_2}
			+ \g \,\phim_{k_1 -k_2} \,,\\ \\
		\left[G^{< >}_{ k_1 k_2 } \right]^{-1} & = & 
				\g \,\phim_{k_1 -k_2} \,,\\ \\
		\left[G^{> <}_{k_1 k_2} \right]^{-1}   & = & 
				\g \,\phim_{k_1 -k_2}\,, \\ \\
		\end{array} 
		\end{equation}
		where we use the notation
		\begin{equation} \label{eq:def-Phi-matrix} 
		\begin{array}{rcl}
		\phim_{k_1 -k_2}~=~\dfrac{1}{L} \left[\pauli~\cdot~\vecphi_{k_1 -k_2}\right] \Theta_{q^\ast -|\veck_1-\veck_2 \mp \vecafm|}\,.
		\end{array} 
		\end{equation}
		The fermionic momentum summations
		are restricted to the vicinity of the hot spots by the step function.
		The dynamics of the low-energy sector and the high-energy sector are described by $[G^{< <}_{ k_1 k_2} ]^{-1}$ and $[G^{> >}_{ k_1 k_2} ]^{-1}$, respectively.
		The low-energy sector is identical to the original spin-fermion model in \eqref{eq:action_spin-fermion-model-mom}.
		In the high-energy sector, the free-fermionic Green's functions $G^{>}_{0,k}$ are shifted by the energy $E_F$ relatively to the Green's function of the low-energy sector $G^{<}_{0,k_1}$.
		The non-diagonal elements $[G^{< >}_{ k_1 k_2 }]^{-1}$ and $[G^{> <}_{k_1 k_2}]^{-1}$ describe transitions of fermions between the low-energy sector and the high-energy sector due to scattering by spin density fluctuations.
		
		In order to construct an effective low-energy theory which deals with low-energy fermions $\Psi^{<}_{k}$ only, we integrate out the high-energy fields $\Psi^{>}_{k}$ from the functional field integral of the partition function $\Z$:
		\begin{equation} \label{eq:partition_function} 
		\begin{array}{rcl}
		\Z 			& = & 		\myint\D(\Psiq^{<}\Psi^{<}) \myint\D (\Psiq^{>}\Psi^{>}) \myint\D(\vecphi^\ast\vecphi) \;
								\e^{ -\action } \\
					& = & 		\myint\D(\Psiq^{<}\Psi^{<}) \myint\D(\vecphi^\ast\vecphi) \;
								\e^{ -\action_{\eff} } \,.
		\end{array} 
		\end{equation}
		The occurring Gaussian integration over the Grassmann fields $\Psi^{>}_{k}$ is an exact transformation without loss of dynamical information.
		The effective low-energy action is the sum of three contributions:
		\begin{equation} \label{eq:action_eff} 
		\begin{array}{rcl} 				
		\action_{\eff} 	& = & 		\action^{<} + \action_{\phi\phi} + \action^{<}_{\Psi\phi\phi}\,.
		\end{array} 
		\end{equation}
		$\action^{<}$ repeats the original spin-fermion action of \eqref{eq:action_spin-fermion-model-mom} in terms of low-energy fields $\Psi^{<}_{k}$.
		The additional contributions $\action_{\phi\phi}$ and $\action^{<}_{\Psi\phi\phi}$ result from the integration procedure.
		The first additional contribution,
		\begin{equation} \label{eq:action-tr-log} 
		\begin{array}{rcl} 				
		\action_{\phi\phi}	& = & -\tr \ln \left[ G^{> >} \right]^{-1} , \\ \\
%							%
%							%& = & 	-\mysum_{k\sigma}\ln\left( \beta G^{>}_{0,k}\right)  \;
%							%		-\mysum_{n=1}^{\infty} \dfrac{(-1)^n}{n} \dfrac{\g^n}{\beta^n} \tr\left\{ \hat{G}^{>}_{0} \left[ \pauli \cdot \vecphi\right]\right\} \\ \\
%							%
%							& = & 	-\mysum_{k\sigma}\ln\left( \beta G^{>}_{0,k}\right) \\
%							&   &	-\dfrac{\gsq}{2 \beta^2 L^2} \mysum_{q k} G^{>}_{0,k} G^{>}_{0,k+q+\vecafm} \left| \vecphi_{q} \right|^2
%									%\vecphi_{-q} \cdot \vecphi_{q} \;
%									+ ... \\ \\
		\end{array} 
		\end{equation}
		contains the action of non-interacting high-energy fermions and, furthermore, describes the impact of the high-energy fermions on the dynamics of the spin density fluctuations.
		The high-energy Green's function provides the inverse energy scale $1/E_F$.
		For weak spin-fermion coupling, the ratio $\gsq/E_F$ is a small parameter and the dynamical contribution of $\action_{\phi\phi}$ to the spin susceptibility is suppressed by  $\gsq/E_F \ll 1$.
		The second additional contribution, 
		\begin{equation} \label{eq:action-int} 
		\begin{array}{rcl} 				
		\action^{<}_{\Psi\phi\phi}	%& = & 	-\mysum_{k_1 k_2}   
										%		\overline{b}^{>}_{k_1}  G^{> >}_{k_1 k_2}  b^{>}_{k_2}\\ \\
										%& = & - \dfrac{1}{\beta^2} \mysum_{k_1 k_2 k_3 k_4}
										%		\overline{\Psi}^{<}_{k_1} \left[ G^{< >}_{k_1 k_2} \right]^{-1}
										%		G^{> >}_{k_2 k_3}
										%		\left[ G^{> <}_{k_3 k_4} \right]^{-1} \Psi^{<}_{k_4} \\

%										& = & - \dfrac{1}{\beta^2} \mysum_{\{k_i\}} %\mysum_{k_1 k_2 k_3 k_4}
%												\overline{\Psi}^{<}_{k_1} \left[G^{< >}_{ k_1 k_2 } \right]^{-1}
%												G^{> >}_{k_2 k_3}
%												\left[G^{> <}_{ k_3 k_4 } \right]^{-1} \Psi^{<}_{k_4} \\
%												%
%										& = & - \dfrac{\gsq}{\beta^2} \mysum_{\{k_i\}} %\mysum_{k_1 k_2 k_3 k_4}
%												\overline{\Psi}^{<}_{k_1} \left[ \pauli \cdot \vecphi_{k_1 - k_2} \right]
%												G^{> >}_{k_2 k_3}
%												\left[ \pauli \cdot \vecphi_{k_3 - k_4} \right] \Psi^{<}_{k_4} \\
										& = & - \dfrac{\gsq}{\beta^2} \mysum_{\{k_i\}} %\mysum_{\small \begin{array}{c} k_1 k_2 \\ k_3 k_4 \end{array} }
												\overline{\Psi}^{<}_{k_1}
												\phim_{k_1 - k_2}
												G^{> >}_{k_2 k_3}
												\phim_{k_3 - k_4}
												\Psi^{<}_{k_4} \,,
		\end{array} 
		\end{equation}
		introduces a new interaction vertex with respect to the low-energy fermions.
		$G^{> >}$ denotes the inverse matrix of $[G^{> >}]^{-1}$, defined in \eqref{eq:critical_action_<_>_matrix_elements}.
		The structure of the interaction vertex can be interpreted as successive scattering of fermions by spin density fluctuations involving high-energy states.
		A fermion in an initial low-energy state is coupled to a first spin density fluctuation and accesses high-energy degrees of freedom described by the matrix $G^{> >}$. 
		Subsequently, the coupling to a second spin density fluctuation scatters the fermion to a final low-energy state.
		
		The representation $\action_{\eff}$ is the desired critical low-energy theory as it contains low-energy fermionic excitations only.
		The extended action $\action$ describes the dynamics of the system in terms of an interaction between the low- and high-energy fermions with coupling constant $\g$ explicitly,
		whereas $\action_{\mathrm{eff}}$ describes the dynamics of the system in terms of low-energy fermions exposed to an effective interaction with coupling constant $\gsq$ and the interaction vertex $\phim G^{> >} \phim$.
		In other words, the effective low-energy action $\action_{\eff}$ is derived at the expense of a more complex interaction vertex between the low-energy fermions.
		
		We cannot treat the effective action exactly and restrict ourselves to the weak coupling limit.
		For weak spin-fermion coupling, the action $\action_{\mathrm{eff}}$ in \eqref{eq:action-int} is suitable to an expansion in the small ratio $\gsq/E_F$.
		Furthermore, the deviations of bosonic momenta $q^\ast$ are small compared with the AFM ordering wave vectors $\vecafm$.
		We only consider the lowest orders in the expansion of $\gsq/E_F$ and $q^\ast/|\vecafm|$.
			The matrix elements of $G^{> >}$ are determined by the equation 
			\begin{equation} \label{eq:matrix_inversion} 
			\begin{array}{rcl}
			\left[ G \right]^{> >}_{k_1 k_2} 
			& = & 
			\dfrac{1}{\beta} G^{>}_{0,k_1} \updelta_{k_1 k_2} \\
			&   &
			\qquad -\dfrac{\g}{\beta}\,G^{>}_{0,k_1} \mysum_{k_3} \phim_{k_1 -k_3} \left[ G \right]^{> >}_{k_3 k_2}\,.
			\end{array} 
			\end{equation}
			The free Green's function of high-energy fermions $G^{>}_{0,k}$ contributes a factor of $1/E_F$ such that $\gsq\,G^{>}_{0,k}$ introduces the small parameter $\gsq/E_F$. 
			The matrix $G^{> >}$ can be calculated to arbitrary accuracy by iterating \eqref{eq:matrix_inversion}.
			In this paper, we consider the leading order in $\gsq/E_F$,
			\ie we approximate $G^{> >}$ by the inverse of the diagonal part
			\begin{equation} \label{eq:matrix_inversion_leading_order} 
			\begin{array}{rcl}
			G ^{> >}_{k_1 k_2} & \approx & 
			\dfrac{1}{\beta} G^{>}_{0,k_1} \updelta_{k_1 k_2} \,.
			\end{array} 
			\end{equation}
			Inserting the approximation of \eqref{eq:matrix_inversion_leading_order} into \eqref{eq:action-int}, the interaction contribution $\action^{<}_{\Psi\phi\phi}$ reads as
			\begin{equation} \label{eq:interaction_leading_order} 
			\begin{array}{rcl}
			\action^{<}_{\Psi\phi\phi} 	& = & 
			- \dfrac{\gsq}{\beta^3}\mysum_{k_1 k_2 k_3}
			\Psiq^{<}_{k_1}
			\phim_{k_1 -k_3}
			G^{>}_{0,k_3}
			\phim_{k_3 -k_2}
			\Psi^{<}_{k_2} \\ \\
			%
%			& = & - \dfrac{\gsq}{\beta^3 L^2}\mysum_{k} \mysum_{q q_1}
%			\Psiq^{<}_{k+q+q_1}
%			\left[ \pauli\cdot \vecphi_{q-\vecafm} \right]
%			G^{>}_{0,k + q_1 +\vecafm}
%			\left[ \pauli\cdot \vecphi_{q_1+\vecafm}\right]
%			\Psi^{<}_{k} \\ \\
			%
			& = & 
			- \gsq \myintl_{x, x^\prime}
			\Psiq^{<}_x
			\phim^\ast_x
			G^{>}_{0,x-x^\prime}
			\phim_{x^\prime}
			\Psi^{<}_{x^\prime} \,.\\ \\
			\end{array} 
			\end{equation}
			The momenta $\veck_1$ and $\veck_2$ are restricted to the Fermi shell while $\veck_3$ is off shell.
			It is convenient to turn back to the coordinate representation $x=(\vecr,\tau)$.
			
			The field $\Phi_x~\equiv~\pauli~\cdot~\vecphi_x$ corresponds to the scattering of a fermion to a high-energy state while the complex-conjugate field corresponds to the scattering from the high-energy state back to the Fermi surface.
			The matrix product $\phim^\ast_x \phim_{x^\prime}$ divides the interaction vertex into a scalar channel and a spin-dependent channel:
			\begin{equation} \label{eq:sigma_matrix_product} 
			\begin{array}{rcl}
			[ \pauli\cdot \vecphi^\ast_x] [ \pauli\cdot \vecphi_{x^\prime}] 
			& = & 
			\vecphi^\ast_x \cdot \vecphi_{x^\prime} \id
			+ \I \pauli \left(  \vecphi^\ast_x \times \vecphi_{x^\prime}\right) \,.
			\end{array} 
			\end{equation}
			Equation~\eqref{eq:sigma_matrix_product} also reveals that the spin-dependent channel only contributes for a non-local interaction.
			Approximating the high-energy Green's function by the inverse high-energy scale $G^{>}_{0,k}\approx 1/E_F$ renders the high-energy Green's function a structureless constant.
			In coordinate space, this approximation corresponds to
			\begin{equation} \label{eq:locality-approx} 
			\begin{array}{rcl}
			G^{>}_{0,x-x^\prime} \approx \dfrac{1}{E_F}\updelta_{x-x^\prime}\,.
			\end{array} 
			\end{equation}
			Therefore, the interaction vertex in \eqref{eq:interaction_leading_order} becomes local, instantaneous, and the spin-dependent channel of the interaction vanishes.
			The interaction vertex takes the form of a local interaction vertex between low-energy fermions and spin density fluctuations with effective coupling constant $\gsq/E_F$:
			\begin{equation} \label{eq:interaction_leading_order_local} 
			\begin{array}{rcl}
			\action^{<}_{\Psi\phi\phi}
			& = & 
			- \dfrac{\gsq}{E_F} 
			\myintl_{x}
			\Psiq^{<}_x\Psi^{<}_x \,\vecphi^\ast_x \cdot \vecphi_x  \\ \\
%			& = & 
%			- \dfrac{\gsq}{E_F} 
%			\mysum_{a}
%			\myintl_{x}
%			\Psiq^{<}(x)\Psi^{<}(x) \,\vecphi^{a}_{\slow}(x) \cdot \vecphi^{a}_{\slow}(x)  \\ \\
%			%
%%			- \dfrac{\lambda}{\beta^2}\displaystyle\sum\limits_{n_1 n_2} \displaystyle\sum\limits_{\veck_1 \veck_2} \displaystyle\sum\limits_{\sigma_1}
%%						\overline{\Psi}^{<}_{\veck_1 \sigma_1 n_1} \left[  \vecphi\cdot \vecphi \right]_{\veck_1 -\veck_2, n_1 -n_2} \updelta_{\sigma_1 \sigma_2} \Psi^{<}_{\veck_2 \sigma_2 n_2} \\ \\
			\end{array} 
			\end{equation}
			In diagrammatic language, integrating out fermionic high-energy fields ties two wavy lines of spin density fluctuations to the same spin-fermion vertex (see Ref.~\onlinecite{HAR11}).
			In Fig.~\ref{fig:Perturbation-series-modified}, the integration procedure corresponds to the first arrow.
			Using the usual logic of a renormalization group analysis, \eqref{eq:interaction_leading_order_local} corresponds to a new interaction, generated by high-energy processes.
			By power counting, this new interaction is less relevant than the leading $\Psiq_x \pauli \Psi_x\cdot\vecphi_x$ term and is usually neglected.
			While this is correct if one is interested in the thermodynamic behavior, it turns out that the interaction in \eqref{eq:interaction_leading_order_local} gives rise to singular corrections to the resistivity of a weakly disordered metal.

	\subsubsection{Effective fermion-fermion interaction} \label{sec:perturbation-lambda-GF}
\begin{figure}[t]
	\centering
	\includegraphics[width=8.6cm]{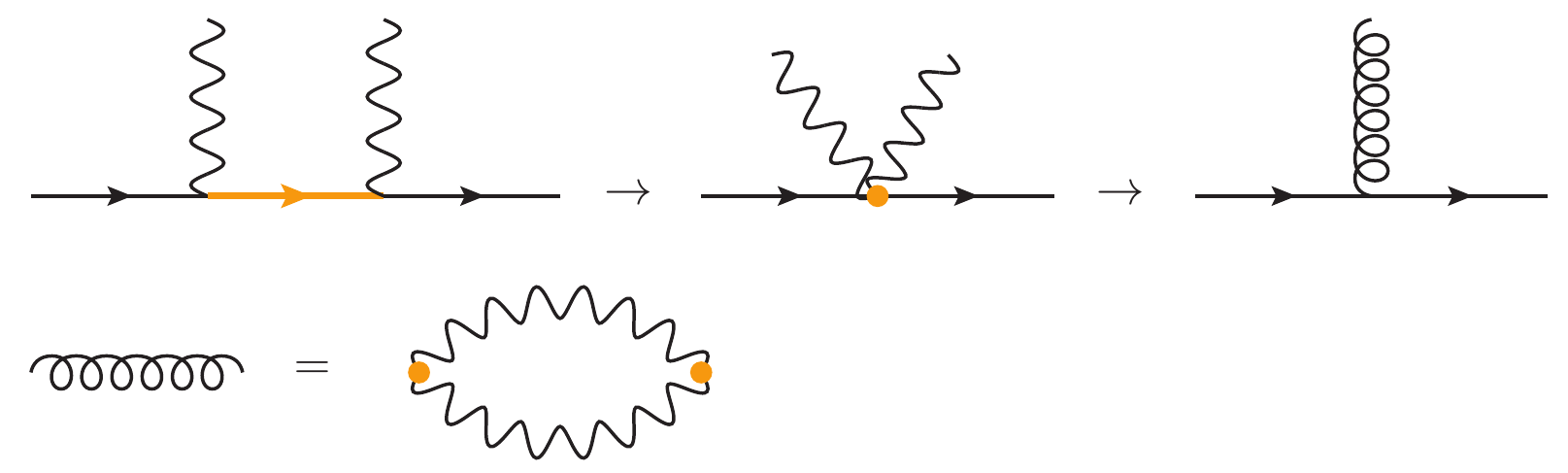}
	\caption{Illustration of the integration procedure:
	In the first step, two spin density fluctuations $\vecphi$ are tied together to a scalar vertex $\vecphi \cdot \vecphi$.
	The original high-energy fermion line (orange line) is approximated by $1/E_F$ (orange dot).
	The composite spin-fermion vertex translates to a fermion-fermion interaction described by a composite propagator, represented by the curly line.
	The composite propagator $\C_{x-x^\prime} \sim \chi^2_{x-x^\prime}$ combines two propagators of spin density fluctuations $\chi_{x-x^\prime}$, represented by the wavy lines.
	}
	\label{fig:Perturbation-series-modified}
	\end{figure}

	Finally, we integrate out the order parameter field $\vecphi$ in order to obtain the effective fermion-fermion interaction $\action_{\ian}$ which is induced by the second-order scattering process:
	\begin{equation} \label{eq:} 
		\begin{array}{rcl}
		\Z 	& = & \myint \D(\Psiq^{<}\Psi^{<})\,\D(\vecphi^\ast\vecphi) \, \e^{-\action_{\phi}-\action_{\Psi\phi\phi}^{<}} \\ 
			& = & \myint \D(\Psiq^{<}\Psi^{<}) \e^{-\action_{\ian}} \,.
		\end{array} 
		\end{equation}
	We expand the interaction contribution with respect to $\gsq/E_F$.
	The first order in $\gsq/E_F$ is an inessential shift of the fermion spectrum.
	At order $\gqu/E_F^2$, the composite vertex translates into the fermion-fermion interaction vertex
	\begin{equation} \label{eq:effective-int-comp} 
		\begin{array}{rcl}
		\action_{\ian} & = & \dfrac{1}{2\beta^3 L^2}\mysum_{k_1 k_2 q} \C_{q} \Psiq_{k_1 +q} \Psiq_{k_2 - q} \Psi_{k_2} \Psi_{k_1}\,.
		\end{array} 
		\end{equation}
	The effective interaction propagator is the convolution of spin susceptibilities which we refer to as composite propagator:
	\begin{equation} \label{eq:composite_propagator} 
	\begin{array}{rcl}
	\C_q 	& = & 	\dfrac{3 \Nhs \gqu}{E_F^2} \,
				\dfrac{1}{\beta L^2}\mysum_{q_1}
				\chi_{q_1} \chi_{q-q_1} \,. \\ \\
	\end{array} 
	\end{equation}
	The factor of 3 results from the three independent polarizations of spin density fluctuations.
	The summation over linearly independent AFM ordering vectors yields the factor of $\Nhs$.
	These vectors are equivalent with respect to adding a reciprocal lattice vector.
	The integration of the order-parameter field corresponds to the transition from the second to the third diagram in Fig.~\ref{fig:Perturbation-series-modified}.
	The composite propagator is represented by the curly line.
	
	For weak spin-fermion coupling $\gsq/E_F \ll 1$, the composite propagator effectively mediates the composite interaction of \eqref{eq:interaction_leading_order_local} and determines the dynamics of the fermions.
	As both momenta $\vecq_1$ and $\vecq -\vecq_1$ are small deviations from $\pm\vecafm$ and restricted by $q^\ast$, the total momentum transfer of a composite mode is small.
	This implies that the composite propagator transfers small momenta.
	Fermions on the entire Fermi surface are subject to the new interaction.
	
		For a disordered system, the question arises as to whether impurity scattering modifies the composite propagator, \ie whether the composite propagator should be averaged with respect to disorder (\cf Sec.~\ref{sec:imp-model}).
		In order to answer this question, we again consider the self-energy correction of the fermionic Green's function in Fig.~\ref{fig:idea-composite-propagator}.
		In principle, impurity scattering involves the intermediate high-energy fermions as well as the low-energy states.
		However, the lifetime $1/E_F$ of the high-energy states is shorter than the elastic scattering time $\tau$ if we apply the main approximation $ E_F\tau \gg 1$.
		As a result, impurity scattering is not relevant during the successive scattering via intermediate high-energy states~\cite{SYZ-notes}
		and we are able to join the spin susceptibilities to the composite propagator of \eqref{eq:composite_propagator}.
		Of course, the impurity scattering of \emph{low-energy} fermions must be analyzed with great care.

	\subsection{Evaluation of the composite propagator} \label{sec:comp-prop-eval}
	In this section, we evaluate the composite propagator on the real axis at $T=0$.
	For $d=2$ dimensions the composite propagator takes the form
	\begin{eqnarray} \label{eq:composite_propagator_retarded} 
	\C^R(\vecq,\omega) 	& = & \re \C^R_0\left(\tilde{q}/q \right) 
+ \re \delta \C^R\left(\gamma |\omega|/(2 q^2) \right) 
\nonumber\\							
&& 
\nonumber\\
&&
+ \I\im \C^R \left(\gamma |\omega|/(2 q^2) \right)\,,
	\end{eqnarray}
	with the momentum cutoff $\tilde{q}^2=\gamma\cutoffe/2$.
	
	The imaginary part of the retarded composite propagator (in $d$ dimensions) was found in Ref.~\onlinecite{ABRa14}.
	We adopt this result,
	\begin{equation} \label{eq:composite_propagator_im_part} 
	\begin{array}{rcl}
	\im \C^R(\vecq,\omega) 	& = & 	\dfrac{3\Nhs\gqu}{E_F^2}\, \myint \dfrac{\dd^2 q_1}{(2\pi)^2} 
										\myintl_{0}^{\omega}\dfrac{\dd\Omega}{2\pi} \\ \\
								&   & \; \times \im\chi^{\mathrm{R}}_{\vecq_1}(\Omega) \im\chi^{\mathrm{R}}_{\vecq-\vecq_1}(\omega -\Omega) \\ \\
								& \approx & \dfrac{\lambda}{\nuf} \sign(\omega) \dfrac{\pi}{2}\dfrac{ \left(\frac{|\omega|\gamma}{2 q^2} \right)^{\frac{3}{2}}}{\left(1 +\frac{|\omega|\gamma}{2 q^2}\right)^{\frac{3}{2}} } \,, \\ \\
	\end{array} 
	\end{equation}
	and introduce the effective dimensionless coupling constant of the composite interaction
	\begin{equation} \label{eq:eff-const} 
	\begin{array}{rcl}
	\lambda & = & \dfrac{2 \Nhs}{\pi^3} \,\dfrac{\gqu}{E_F^2} \, \dfrac{\nuf}{\gamma}\,,
	\end{array} 
	\end{equation}
	where $\nuf$ denotes the fermionic density of states at the Fermi level. 
	The imaginary part is a scale-invariant function, \ie it only depends on the ratio $\omega/q^2$.
	The integrals are performed by employing the convolution theorem and by means of interpolating functions~\cite{WOE-notes}.
	In the calculation, the overdamped form of the spin susceptibility is used and the dynamical contribution $\sim \omega^2$ in \eqref{eq:spin-sus-form} is neglected,
	\ie we assume that the damping term $\I \gamma \omega$ is larger than the dynamical contribution $\omega^2 /c^2$ for the relevant frequencies.
	Since the imaginary part of the retarded spin propagators 
	decays as $1/q^4$ for large $q$, the cutoff $q^\ast$ is lifted from the momentum integral.

	The calculation of the interaction correction to the Drude conductivity requires both the real and the imaginary parts of the composite propagator [\cf \eqref{eq:sigma1_sigma2_sigma3_averaged}].
	In order to find the real part, we apply Kramers-Kronig relation~\cite{ELK79} to the composite propagator,
	
		\begin{equation} \label{eq:composite_propagator_re_part_KK} 
		\begin{array}{rcl}
		\re C^R(\vecq,\omega) 	& = & \dfrac{1}{\pi}\mathcal{P}\ds\int\limits_{-\infty}^{\infty} \dd\omega^\prime \dfrac{\im C^R(\vecq,\omega^\prime)}{\omega^\prime -\omega} \,.\\ \\
%									& = & \dfrac{2}{\pi}\mathcal{P}\ds\int\limits_{0}^{\infty} \dd\omega^\prime \im C^R(\vecq,\omega^\prime)\dfrac{\omega^\prime}{{\omega^\prime}^2 -\omega^2} \\ \\
		\end{array} 
		\end{equation}
		The integral is formally UV divergent.
		This is a consequence of the overdamped form of the spin susceptibility which is inserted in \eqref{eq:composite_propagator_im_part}.
		Therefore, we need to cut off the frequency integral at the scale $\cutoffe$.
		In doing so, the integral depends on two parameters: the cutoff energy and the external frequency. 
		The cutoff dependence can be separated from the $\omega$-dependent part by subtracting the zero-frequency contribution ${\re C^R(\vecq,\omega~=~0)~\equiv~\re \C^R_0\left(\tilde{q}/q\right)}$.
		The integral of the cutoff-dependent part yields
		\begin{equation} \label{eq:composite_propagator_re_part_0} 
		\begin{array}{rclcl}
		\re \C^R_0 \left(\tilde{q}/q \right)	& = & 
											\dfrac{2}{\pi}\mathcal{P}\ds\int\limits_{0}^{\cutoffe} \dd\omega^\prime \dfrac{\im C^R(\vecq,\omega^\prime)}{\omega^\prime}  \\ \\
%									& = & 
%											 \dfrac{\lambda}{\nuf} \mathcal{P}\ds\int\limits_{0}^{\tilde{q}^2/q^2} \dd u
%												 \dfrac{u^{\frac{3}{2}}}{\left(1+u\right)^{\frac{3}{2}} } \dfrac{1}{u} \\ \\
%									& = & 
%										 \dfrac{\lambda}{\nuf}\mathcal{P}\ds\int\limits_{0}^{\tilde{q}/q} \dd x
%												 \dfrac{2 x^2}{\left(1+x^2\right)^{\frac{3}{2}} }
%									& \approx & 
%										 \dfrac{\lambda}{\nuf}\mathcal{P}\ds\int\limits_{0}^{\tilde{q}/q} \dd x
%												 \dfrac{2 x^2}{1+x^3 } \\ \\
									& \approx &  \dfrac{\lambda}{\nuf}\,\dfrac{2}{3} \ln\left[ 1 + \left(\dfrac{\tilde{q}}{q}\right)^3 \right]\,.
		\end{array} 
		\end{equation}
		
		Here the $\omega$-dependent part 
$\re \delta \C^R \left(\gamma |\omega|/(2 q^2) \right)\linebreak =\re \C^R(\vecq,\omega)-\re \C^R(\vecq,\omega=0)$ 
is convergent, hence the cutoff restriction can be lifted.
		Then, ${\re \delta \C^R}$ is a scale invariant function.
		In the limit of small and large ratios ${x=\gamma|\omega|/(2q^2)}$, the asymptotic forms read as
		\begin{equation} \label{eq:composite_propagator_re_part_w_res-asym} 
		\begin{array}{rcl}
\re \delta \C^R(x \ll 1)	& \simeq & \dfrac{\lambda}{\nuf} \dfrac{\pi}{2}\,x^{\frac{3}{2}} \,,\\ \\
\re \delta \C^R(x \gg 1)	& \simeq & -\dfrac{\lambda}{\nuf} \ln(x) \,.\\ \\
		\end{array} 
		\end{equation}

		In the static limit,
		${\omega =0}$
		the scale invariant functions are zero and the composite propagator exhibits a logarithmic singularity due to the cutoff-dependent contribution.
		For ${x=\gamma |\omega|/(2 q^2) \gg 1}$ the composite propagator takes the asymptotic form:
		\begin{equation} \label{eq:composite_propagator_wgg} 
		\begin{array}{rcl}
		\C^{R,A}(\vecq,\omega) & \simeq & \dfrac{\lambda}{\nuf} \left[
										\ln\left( \dfrac{\cutoffe}{|\omega|} \right) \pm  \dfrac{\I\pi}{2} \sign(\omega)
										\right]\,.
		\end{array} 
		\end{equation}
		The positive and the negative signs refer to the retarded and the advanced propagators, respectively.
		The cutoff-dependent $\ln(q)$~contribution
		cancels the scale-invariant contribution of ${\re \delta \C^R}$.
		The logarithmic singularity cannot be avoided irrespective of the order of the limits  $\omega \rightarrow 0$ and $q \rightarrow 0$.
		
		For a disordered system, the question arises as to whether the logarithmic singularity of the composite propagator conspires with the diffusion pole and produces singular interaction corrections to Drude conductivity, similar to the Coulomb interaction.

\section{Impurity model} 
\label{sec:imp-model}

	At low temperature, the finite conductivity of metals is attributed to elastic scattering of electrons and static impurities.
	Our treatment of impurity scattering relies on a statistical approach:
	Under the assumption that the phase relaxation length of the electrons $\lphi$ is much smaller than the system size $L$,
	a macroscopic system of size can be regarded as network of statistically independent subsystems of size $\lphi$.
	Then, the conductivity of the total system is the average over a large number of subsystems or the average over all impurity configurations, \ie the impurity positions are treated as random quantities.
	A measured observable of the disordered electron system is the impurity average of the respective physical quantity.
	The diagrammatic rules of the impurity averaging are developed by expanding the observable of interest in the total impurity potential $U(\vecr)$ and by averaging each term of the perturbation series~\cite{ABRb63,ALTb85}.
	For the random impurity potential we assume Gaussian white-noise disorder with
	\begin{equation} \label{eq:imp-pot-correl}	
	\begin{array}{rcl}
	\Braket{U(\vecr)} 					& = & 0 \,,\\
	\Braket{U(\vecr) U(\vecr\,^\prime)} & = & \dfrac{1}{2\pi \nuf \tau} \updelta(\vecr -\vecr\,^\prime) \,,
	\end{array} 
	\end{equation}
	where $\tau$ is the elastic scattering time.
	In momentum space, the two-point correlation corresponds to a structureless impurity line which transfers momentum only.
	Furthermore, we assume a ``good metal,'' \ie
	\begin{equation} \label{eq:main-app}	
	\begin{array}{rclcrcl}
	\cond	 & \gg & 1  & \text{or} &
	%\drude	 & \gg & \dfrac{e^2}{\hbar} & \text{or} &
	E_F \tau & \gg & 1 \,.
	\end{array} 
	\end{equation}	
	%%%% CHANGES
	Here, ${\cond = \drude \hbar/ e^2}$ denotes the dimensionless conductance.
	%%%%
	Within this main approximation, diagrams with crossing impurity lines are suppressed by factors of ${1/E_F \tau \ll 1}$ and we are able to perform a controlled expansion in this parameter.
	
	Observables of the electron system are typically expressed as impurity-averaged combinations of Green's functions.
	In particular, the building block of the impurity-averaged Green's function
	\begin{equation} \label{eq:averaged-GF}
	\begin{array}{rcl}
	\Braket{ G^{R/A} (\mathbf{p},\epsilon)} 	& = & \dfrac{1}{\epsilon -\epsilon_{\mathbf{p}} \pm \frac{\I}{2\tau} }
	\end{array} 
	\end{equation}
	appears as building block.
	The impurity average of two Green's functions together introduces correlations between two propagating electrons which are represented by impurity lines connecting the formerly separate fermion lines.
	The soft modes of the impurity-induced vertex are the diffuson and the cooperon channel.
	The diffuson channel is the geometric series of ladderlike diagrams with each number of impurity lines.
	Summing up the geometric series yields~\cite{ZNA01}
	\begin{equation} \label{eq:diffuson_series}
	\begin{array}{rclcl}
			\D(\vecq,\Omega) 
			& = &
			\dfrac{1}{2\pi\nuf\tau} \dfrac{S}{S-\frac{1}{\tau}} \,,
	\end{array} 
	\end{equation}
 	where the function $S$ is defined:
	\begin{equation} \label{eq:function-S}
	\begin{array}{rcl}
	S(\vecq,\Omega)  & = & \sqrt{ \left( \I\Omega +\dfrac{1}{\tau} \right)^2 +v_F^2 q^2 }
	\end{array} 
	\end{equation}
	The diffuson channel is the relevant soft mode for small-momentum differences since $\D(\vecq,\Omega)$ exhibits the diffusion pole for small-momentum difference $\vecq$.
	The cooperon channel leads to the weak localization of electrons which we ignore in this paper.

\section{Interaction correction: Composite modes}
\label{int-cor}

For transport in disordered metals we distinguish between the classical Drude conductivity $\sigma_0$ and quantum corrections due to quantum mechanical interference effects in the propagation of the electrons.
The quantum corrections fall into two classes~\cite{ALE99}.
In the one-particle picture, the interferences in the propagation of single electrons lead to the weak localization correction to conductivity~\cite{ALTb85}. 
We do not consider this class of quantum corrections in this paper.
The second class is entirely due to electron-electron interactions beyond the one-particle picture.
These interaction corrections are attributed to the coherent scattering off Friedel oscillations of the electron density around the impurities~\cite{ZNA01}.

For a generic long-range electron-electron interaction the general form of the interaction correction to conductivity was found in Ref.~\onlinecite{ZNA01}.
Since the composite propagator $\C$ exhibits a logarithmic singularity at $q=0$,
this propagator falls into the class of long-range interaction propagators for which the general formula 
is applicable.

\subsection{General form of the interaction correction}
\begin{figure}[t]
		\centering
		\includegraphics[width=8.6cm]{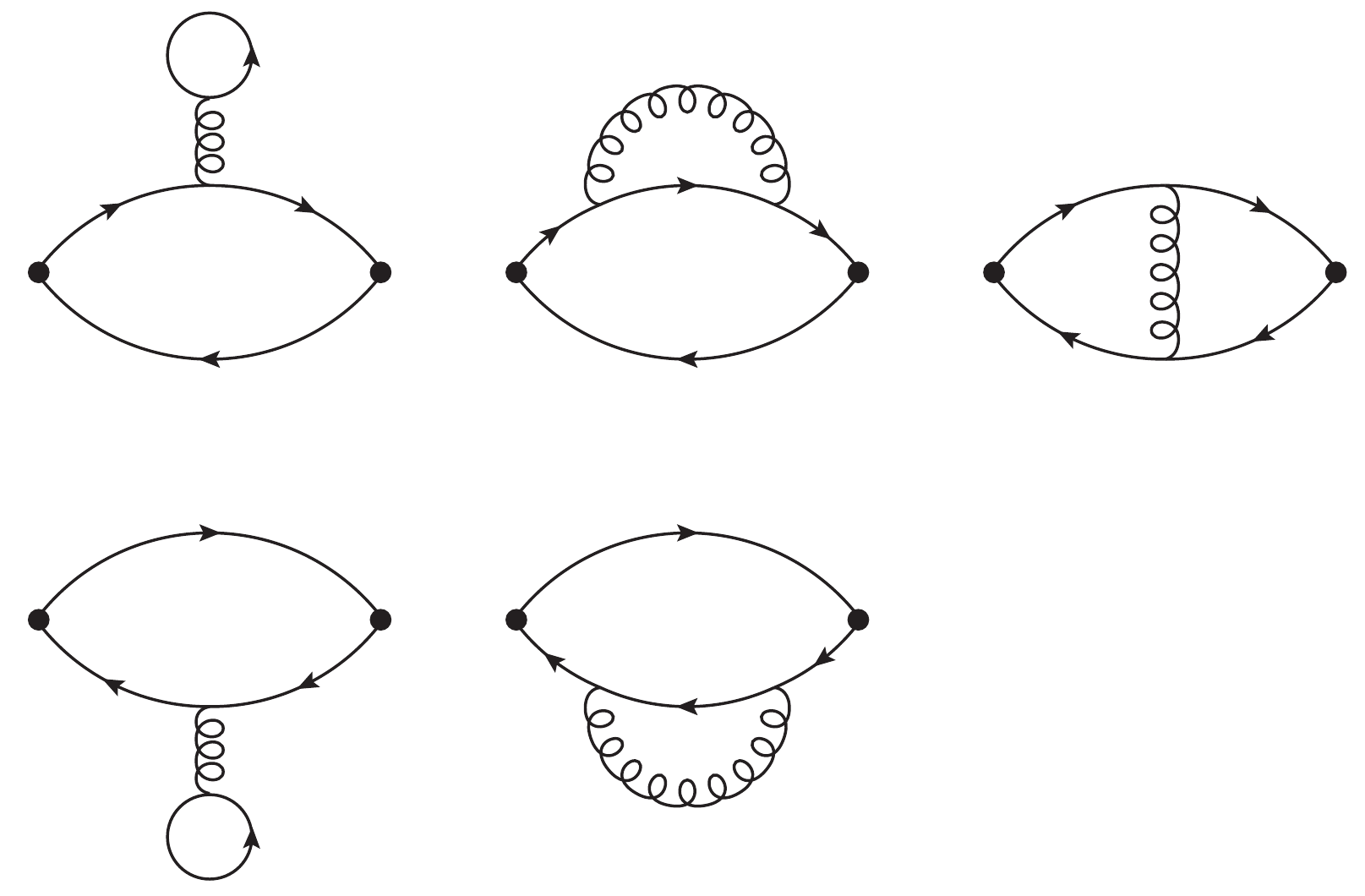}
		\caption{
		Expansion of the current-current correlator to lowest order in the interaction. Solid lines denote fermionic Green's function, curly lines denote composite propagators, black dots denote current operators.
		}
		\label{fig:Current_current_bubble_first_order_relevant}
\end{figure}

Following Ref.~\onlinecite{ZNA01}, the calculation consists of three steps:
(i) Within the linear response, the conductivity is related to the current-current correlator $Q^{\alpha\beta}_{\vecr_1,\vecr_2}(\omega)$ by the Kubo formula
\begin{equation} \label{eq:kubo}	
\begin{array}{rcl}
\sigma^{\alpha\beta}_{\vecr_1,\vecr_2}(\omega) 		
		& = & \dfrac{\I}{\omega} \left[Q^{\alpha\beta}_{\vecr_1,\vecr_2}(\omega) 
%		\right. \\ 
%		&   & \left.
		 + \dfrac{\e^2}{m} n_{\vecr_1}\updelta_{\vecr_1-\vecr_2} \updelta^{\alpha\beta} \right] \,. 
\end{array} 
\end{equation}
The current-current correlator on the imaginary axis $Q^{\alpha\beta}_{\vecr_1,\vecr_2}(\I\omega_n)$ is expanded to the lowest order of the electron-electron interaction, represented by the diagrams in Fig.~\ref{fig:Current_current_bubble_first_order_relevant}.
(ii) After the analytic continuation ${\I\omega_n \rightarrow \omega + \I 0^+}$ of these diagrams, the (real-valued) dc conductivity is obtained by performing the zero-frequency limit
\begin{equation} \label{eq:static}
\begin{array}{rcl}
\sigma^{\alpha\beta}_{\vecr_1,\vecr_2}(\omega=0) 	& \equiv 	& - \lim\limits_{\omega\to 0}\left(\dfrac{\im\,Q^{\alpha\beta}_{\vecr_1,\vecr_2}(\omega) }{\omega}\right).
\end{array} 
\end{equation}
(iii) Finally, the conductivity tensor ${\sigma^{\alpha\beta}_{\vecr_1,\vecr_2}(\omega=0)}$ is averaged with respect to impurity configurations according to the diagrammatic rules of Sec.~\ref{sec:imp-model} and within the main approximation ${1/E_F\tau \ll 1}$.
The impurity-averaging restores the translation invariance of the conductivity tensor.
Our final result is the response to a homogeneous electric field which is related to the impurity-averaged conductivity tensor by spatial integration:
\begin{equation} \label{eq:averaged}
\begin{array}{rcl}
\Braket{\sigma^{x x}(\omega=0) } 	& \equiv	& \displaystyle\int\dd(\vecr_2-\vecr_1) \Braket{\sigma^{x x}_{\vecr_1,\vecr_2}(\omega=0) } \\
\end{array} 
\end{equation}
The impurity-averaged conductivity reflects the isotropy of the system (without a magnetic field), \ie the non-diagonal elements vanish and the diagonal elements are equal.
The interaction correction to conductivity takes the form~\cite{ZNA01}:
\begin{equation} \label{eq:sigma1_sigma2_sigma3_averaged} 
\begin{array}{rcl}
\delta\sigma 	& = & 
										-\dfrac{e^2}{\hbar} v_{F}^2 \nuf \pi
										\myint_{-\cutoffe}^{\cutoffe} \dfrac{\dd \Omega}{4\pi^2} 
										\tf^\prime \left(\frac{\Omega}{T} \right)
										\myint \dfrac{\dd^2 q}{(2\pi)^2} \\
										&   & \times \im\left\{ \Ceff^A(\vecq,\Omega) B(\vecq,\Omega)\right\}	\,,	
\end{array} 
\end{equation}
with the Fermi velocity $v_F$.
The diverging function ${f(x)=x \cth(x/2)}$ requires a frequency cutoff $\cutoffe$.

We include screening of the long-range electron-electron interaction within the random phase approximation.
In the density channel of the electron-electron interaction the effective composite propagator reads as
\begin{equation} \label{eq:subst-V} 
\begin{array}{rcl}
							\Ceff^A(\vecq,\omega)
							& = & 
							\dfrac{\left[\C^A(\vecq,\omega) + F^{\rho}_0\right]}{1 -\left[\C^A(\vecq,\omega) + F^{\rho}_0\right] \Pi^A(\vecq,\omega) } \,,
\end{array} 
\end{equation}
with the Fermi liquid parameter $F^\rho_0$ which we assume to be of order unity or less.

\begin{figure}[t]
		\centering
		\includegraphics[width=8.6cm]{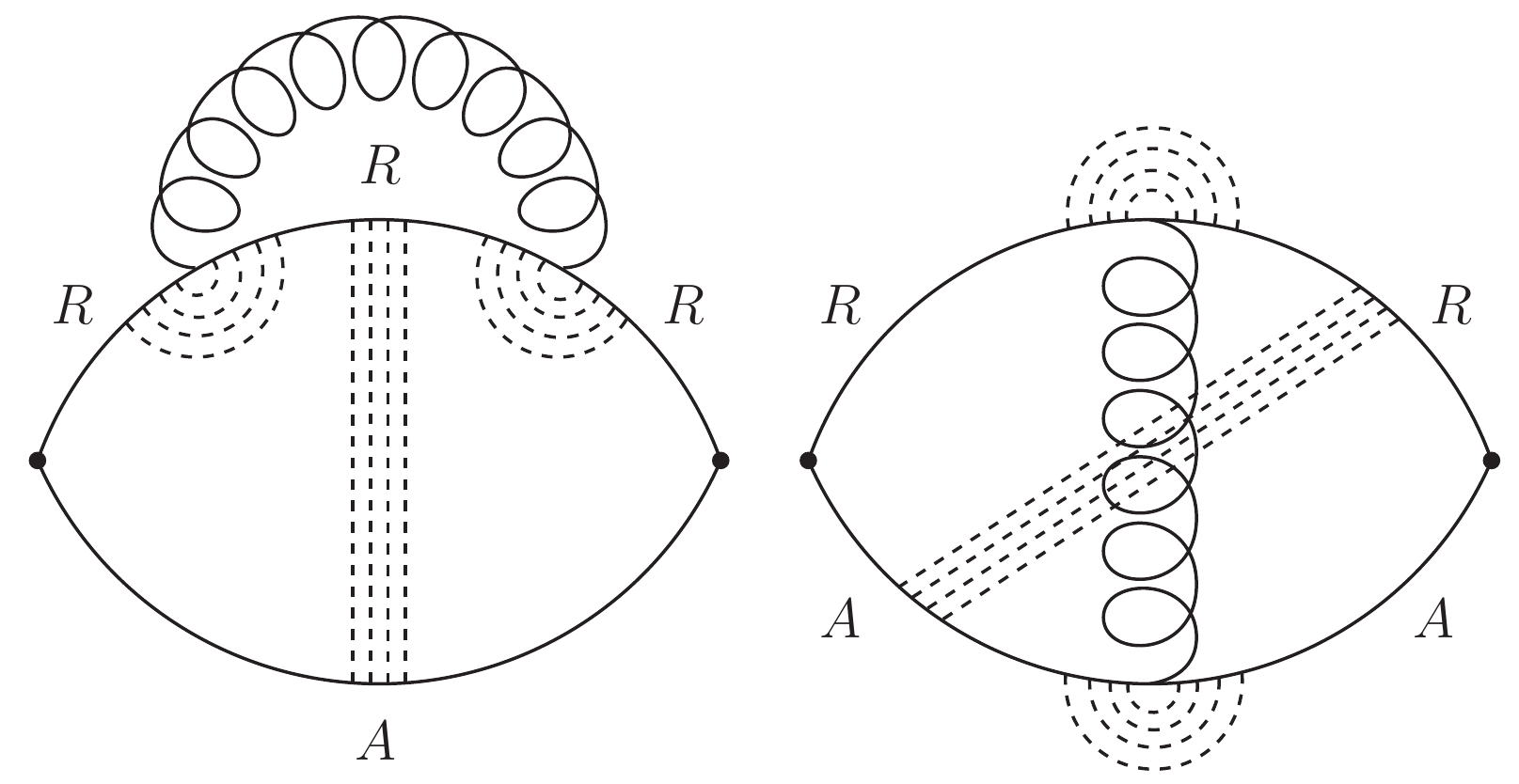}
		\caption{Relevant diagrams of the impurity-averaged current-current correlator in the diffusive limit ${T \tau \ll 1}$.
		${R,A}$ denote retarded and advanced impurity-averaged fermion lines.
		Multiple dashed lines indicate diffuson ladders.
		}
		\label{fig:3rd-order-diffuson-pole}
\end{figure}

We are particularly interested in the interplay between the composite modes and the diffusive modes and we turn to the diffusive limit $T\tau \ll 1$.
In this limit the diagrams of Fig.~\ref{fig:3rd-order-diffuson-pole} with a maximum number of diffuson ladders dominate the correction to conductivity.
Performing the integration over fermionic momenta and real frequencies, the function $B$ and the polarization operator $\Pi^A$ take the explicit forms:
\begin{equation} \label{eq:B-diffusive}
\begin{array}{rcl}
B(\vecq, \Omega) 		& = & \dfrac{2 Dq^2}{\left(Dq^2 + \I\Omega\right)^3} \,, \\ \\
\Pi^A(\vecq, \Omega) 	& = & \dfrac{Dq^2}{Dq^2 + \I\Omega} \,,
\end{array} 
\end{equation}
with the diffusion constant in two dimensions ${D = v_F^2 \tau /2}$.
The 3rd-order diffusion pole of the function $B$ corresponds to the three diffuson ladders of the diagrams in Fig.~\ref{fig:3rd-order-diffuson-pole}.

\subsection{Temperature dependence}
Our goal is to find the asymptotic temperature dependence for low temperatures, \ie for $\cutoffe/T \gg 1$.
For convenience, we introduce the dimensionless integration variables ${\zz=Dq^2/\Omega}$ and ${\xz = \cutoffe/\Omega}$
and rewrite the interaction correction in the diffusive limit as
\begin{equation} \label{eq:ZNA_formula_x_z_diffusive} 
\begin{array}{rcl}
\delta\sigma_{\cp}(T\tau \ll 1) 	& = 	&  -\dfrac{e^2}{\hbar} \dfrac{\nuf}{(2\pi)^2} 
											\im
											\myintl_{0}^{\infty} \dd \zz \tilde{B}(\zz) \\ \\
									&    &	\times \myintl_{1}^{\infty} \dfrac{\dd \xz}{\xz} \tf^\prime\left(\frac{\cutoffe/T}{\xz}\right)\tilde{\C}_{\eff}^A(\xz)\,,\\ \\
\end{array} 
\end{equation}
Since the function ${\tilde{B}(z)= 2z/(\I +z)^3}$ over $z$ decays ${\sim 1/z^2}$ for large $z$, the main contribution of the integral resides in the region of ${z \ll 1}$.
This allows us to approximate the composite propagator under the integral by its asymptotic form
\begin{equation} \label{eq:composite-op-z-ll-1} 
\begin{array}{rcl}
\tilde{\C}^A(\xz) & \simeq & \lambda\left[ \ln \xz - \dfrac{\I\pi}{2} \right] \,.
\end{array} 
\end{equation}
The effective propagator $\tilde{\C}_{\eff}^A(\xz)$ is given by the analog of \eqref{eq:subst-V} where we replace $\Pi^A$ by the dimensionless polarization operator ${\tilde{\Pi}^A(z) = z/(\I + z)}$.

The relevance of screening effects, described by the polarization operator, depends on the balance between thermal energy of the fermions and the interaction energy:
	when the temperature is lowered, the fermions lose kinetic energy.
	They are increasingly influenced by the interaction and redistribute.
	This leads to strong screening of the long-range interaction.
	At lowest temperatures, strong screening renders details of the composite propagator unimportant and the specific interaction is replaced by a universal interaction propagator.
	The universal propagator entails a universal interaction correction for sufficiently low temperatures.
	
	At sufficiently high temperatures, the influence of the interaction is small compared to the kinetic energy of the fermions and the screening is not efficient.
	Above a certain crossover temperature, the screening of the composite propagator can be neglected and the dynamics of the fermions is governed by the bare propagator.
	In this temperature regime, the interaction correction depends on the specific form of the interaction propagator.
	
	The logarithmic form of the composite propagator suggests to introduce the exponentially small crossover temperature
	\begin{equation} \label{eq:T-cross} 
	\begin{array}{rcl}
	T^\ast & = & \cutoffe \e^{-1 /\lambda}\,,
	\end{array} 
	\end{equation}
	which separates the temperature regimes of (i) strong screening, ${T < T^\ast \ll \cutoffe}$, and (ii) weak screening, ${T^\ast < T \ll \cutoffe}$.
	
	(i) For temperatures well below the crossover temperature, ${\lambda \ln\left( \cutoffe /T\right) \gg 1}$, the inverse polarization operator plays the role of the universal propagator ${\tilde{\C}^A = -1/\tilde{\Pi}^A}$.
	Then, the integral yields
	\begin{equation} \label{eq:sigma-below-cross} 
	\begin{array}{rcl}
	\delta\sigma_{\cp}(T<T^\ast \ll \cutoffe) 	
%					& = & 
%					-\dfrac{e^2}{(2 \pi)^2 \hbar} \ln\left( \dfrac{\cutoffe}{T}\right) \, \im \myintl_{0}^{\infty} \dd \zz B(\zz)  \left( -\dfrac{1}{\tilde{\Pi}^A(\zz)}\right) \\ \\
					& = & 
					-\dfrac{e^2}{2\pi^2 \hbar} \ln\left( \dfrac{\cutoffe}{T}\right)\,,
						
	\end{array} 
	\end{equation}
	\ie below the crossover temperature the conductivity is reduced compared to the Drude conductivity. 
	Equation~\eqref{eq:sigma-below-cross} is identical to the interaction correction obtained for the density channel of the Coulomb interaction in the diffusive limit~\cite{ZNA01}.
	
	(ii) For temperatures above the crossover temperature (but still in the asymptotic limit ${\cutoffe/T \gg 1}$) the inequality ${\lambda \ln\xz < \lambda\ln\left( \cutoffe /T \right) \ll 1}$ holds for the frequency variable~$\xz$.
	The unity dominates in the denominator of $\Ceff^A$ and we approximate the effective screened propagator by the bare composite propagator in \eqref{eq:composite-op-z-ll-1}.
	In the asymptotic limit, we find the specific temperature dependence
\begin{equation} \label{eq:sigma-above-cross} 
\begin{array}{rcl}
	\delta\sigma_{\cp}(T^\ast< T \ll \cutoffe) 
%					& = & 
%					-\dfrac{e^2 \lambda}{(2 \pi)^2 \hbar} \left[ \dfrac{1}{2}\ln^2\left( \frac{\cutoffe}{T}\right) + F^\rho_0 \ln\left( \frac{\cutoffe}{T}\right) \right] \, \myintl_{0}^{\infty} \dd \zz \im\{B(\zz)\}   \\ \\
											
%					& = & 
%+\dfrac{e^2\lambda}{(2 \pi)^2 \hbar} \dfrac{1}{2}\ln^2\left( \dfrac{\cutoffe}{T}\right) \\ \\
					& = & 
					+\dfrac{e^2\lambda}{8 \pi^2 \hbar}\ln^2\left( \dfrac{\cutoffe}{T}\right) \,.
						
\end{array} 
\end{equation}
\ie above the crossover temperature, the conductivity is raised compared to the Drude conductivity. 
Remarkably, the interaction correction induced by composite modes is not proportional to powers of the temperature.
	Instead, the interference of composite modes and diffusive modes piles up a strong $\ln^2 T$ correction.
	Furthermore, the interaction correction changes sign from the universal localizing behavior to a specific antilocalizing behavior.

\section{Relevance of composite modes}
\label{disc}

The conductivity for a disordered two-dimensional metal
near an AFM QCP consists of three contributions:
\begin{equation} \label{eq:Drude-conduct-corrections} 
\sigma=\drude + \delta\sigma_{\hh} + \delta\sigma_{\cp}.
\end{equation}
The main contribution is given by the Drude conductivity
of the non-interacting electron system,
${\drude \propto  E_F \tau \gg 1}$.

Two interaction corrections add to the Drude conductivity:
The interaction correction $\delta\sigma_{\hh}$
is induced by AFM spin density fluctuations
which scatter fermions only between hot spots of the Fermi surface.
Adopting the theory of Ref.~\onlinecite{SYZ12}
(which was focusing on the 3D case),
we recalculate this correction
for a two-dimensional metal
which we consider in the present paper.
Under the assumption of Landau damping
the correction can be expressed in terms of the coupling constant $\lambda$, 
\begin{equation} \label{eq:hh-conduct} 
\delta\sigma_{\hh}=- \dfrac{e^2}{\hbar}
			\drude^2
			\dfrac{\pi^3 C(T)}{4\sqrt{3}}\,
			\lambda \,
			\dfrac{T}{E_F}
			\,.
\end{equation}
The function ${ C(T) = \ln\left( \gamma T \xi^{2}\right) }$ depends
weakly on the temperature ($\xi$ denotes the correlation length).

The second interaction correction $\delta\sigma_{\cp}$ is caused by the composite modes.
In situations in which our theory applies,
the effective coupling constant $\lambda$ is small and
the crossover temperature ${ T^\ast \propto \e^{-1/\lambda} }$ is exponentially suppressed.
Then, $\delta\sigma_{\cp}$ is given
by the positive $\ln^2 T$-correction of \eqref{eq:sigma-above-cross}
down to all experimentally accessible temperatures.

The crossover from $\delta\sigma_{\hh}$ to $\delta\sigma_{\cp}$ occurs
at low enough temperatures when $ { \sigma_{\hh}(T) < \delta\sigma_{\cp}(T) }$.
Setting 
${ |\sigma_{\hh}(\tilde{T})| = |\delta\sigma_{\cp}(\tilde{T})| }$, we
establish the second crossover temperature $\tilde{T}$. Our theory
predicts that for $T < \tilde{T}$ the correction $\delta\sigma_{\cp}$
is dominant.

The temperature scale $\tilde{T}$ is determined
by the residual resistivity $\rho_0=\sigma_0^{-1}$ through
\begin{equation} \label{eq:tilde-T} 
\begin{array}{rcrcl}

\dfrac{\tilde{T}}{E_F}
		& = &
			\dfrac{\sqrt{3}}{2  \pi^5 \,C(\tilde{T}) } \,
			\left(\dfrac{\rho_0 e^2}{\hbar} \right)^2 \,
			\ln^2\left( \dfrac{\tilde{T}}{E_F}\right)  
			
			\,,
\end{array} 
\end{equation}
and can be calculated
from experimental data of resistivity measurements at low temperatures.

Using the data of Ref.~\onlinecite{Butch2012} on the electron-doped
cuprate ${ \mathrm{La_{2-x}Ce_{x}CuO} }$ at doping ${ x=0.17 }$ (we
also use the lattice constant $c$ from Ref.~\onlinecite{Sawa2002}), we
find reasonable values of the parameters of our theory. For the upper
crossover scale $\tilde{T}$ below which the $\ln^2T$ correction may
be observable, we find the value ${\tilde{T}\approx1.1}$\,K. At the same
time, the slope of the linear part of the resistivity matches
%$\delta\sigma_{\hh} \sim -\lambda T$ 
\eqref{eq:hh-conduct} with ${\lambda\approx0.033}$, which is
consistent with our previous restriction to the weak-coupling limit.
Therefore, the localizing behavior sets in at very small temperatures
${ T < T^\ast \approx 10^{-9}}$\,K. In contrast, using the data reported
for the somewhat smaller doping level ${x=0.15}$, we find a much
larger value of ${\lambda\approx0.31}$ resulting in a much higher
$T^\ast\approx240$\,K. This suggests a possible breakdown of our
weak-coupling approach at this doping level.

Furthermore, our analysis of the linear resistivity data~\cite{Kasahara2010}
of the iron pnictide
${\mathrm{BaFe_2(As_{1-x}P_x)_2}}$ at doping ${x=0.33}$ yields a large
effective coupling constant ${ \lambda > 1 }$.  This result (together
with the wide range of the linear behavior up to room temperature)
suggests that the linear resistivity in this compound has a different
origin, different from small perturbations by spin fluctuations and
disorder.
Similar conclusions can be drawn for hole-doped cuprate superconductors.

\section{Conclusion}
\label{concl}

In this paper, we have analyzed the interaction correction to
conductivity in a 2D disordered metal near an AFM QCP
with the focus on the interplay between electron scattering off
composite modes of spin density fluctuations and static impurities.
Our study is based on the notion that successive scattering by AFM
spin density fluctuations can, in total, transfer small momenta and
hence can mimic the coherent scattering off Friedel oscillations
leading to an Altshuler-Aronov--type correction to conductivity.

The effective electron-electron interaction mediated by the composite
modes can be found within the low-energy theory of the spin-fermion
model. Unlike the direct exchange of spin fluctuations considered in
Ref.~\onlinecite{SYZ12}, this effective interaction involves processes
of higher order in the spin-fermion coupling. Assuming the
spin-fermion coupling to be weak, we can describe the leading
higher-order process by a composite propagator
(\ref{eq:composite_propagator_wgg}) given by a convolution
(\ref{eq:composite_propagator}) of two spin susceptibilities
(\ref{eq:spin-sus-form}). The real part of the composite propagator
exhibits a logarithmic singularity for small momenta.

Having found the effective interaction capable of small-momentum
transfers over the whole Fermi surface, we followed the standard route~\cite{ZNA01}
to evaluate the corresponding quantum correction to
conductivity. The peculiar form of the effective interaction
(\ref{eq:composite_propagator_wgg}) results in two distinct temperature
regimes. For the lowest temperatures, all features of the composite
propagator are washed out by screening and we recover the
standard localizing correction ${\delta\sigma\propto -\ln T}$, as
expected for generic singlet-channel interactions. However, the
temperature range where screening is effective is limited by the
exponentially small crossover temperature $T^\ast$. Above the
crossover temperature, 
%%%%% CHANGES  
we find a stronger dependence ${\delta\sigma\propto +\ln^2 T}$ (with
a {\it positive} sign).  As a result, the conductivity is a
non-monotonic function of $T$, as sketched in
Fig.~\ref{fig:sigma-plot}. In terms of the original spin-fermion
coupling, the $\ln^2 T$ correction is the second-order contribution
and has to be compared to the first-order result
\eqref{eq:hh-conduct} (see Ref.~\onlinecite{SYZ12}). Given the linear
temperature dependence of \eqref{eq:hh-conduct}, the second-order
correction dominates at temperatures below the crossover scale
$\tilde{T}$. We find that this behavior may be observable in
electron-doped cuprates.

%%%%% CHANGES

An antilocalizing $\ln^2 T$ behavior was previously reported by Kim
and Millis\cite{Kim2003} near a metamagnetic QCP
and by Paul~\textit{et~al.}~\cite{PAU05} near a ferromagnetic QCP.
Our analysis of the corrections near an AFM
QCP shows that the seemingly antilocalizing $\ln^2
T$ correction due to the composite modes appears only in an
intermediate (although wide) temperature regime.  At the lowest
temperatures, ${T<T^*}$, we find the localizing logarithmic behavior,
as expected for a singlet-type interaction.
%
%%%%%%%%%%%
%
A quick estimate shows that
for a three-dimensional metal the quantum correction would scale as
$\sqrt{T}$.

%%%%%% CHANGES

Would interaction corrections to the impurity vertex change our
results? In the absence of disorder, singular backscattering due to
ferromagnetic fluctuations was found in
Refs.~\onlinecite{Kim2003,Rossi2010}. Indeed, in a
clean system the composite modes considered in this paper would also
lead to singular (logarithmic) backscattering. However, disorder
averaging regularizes the vertex function. Technically, the
regularization comes from the finite imaginary part in the Green's
functions \eqref{eq:averaged-GF}. The finite vertex correction can be
absorbed in the definition of $\tau$ and does not lead to any
qualitative changes in the theory. Hence, our theory is applicable to
disordered metals with the finite Drude conductivity $\sigma_0$. In
contrast, we expect our theory to break down if any competing
order should emerge preemptively near the QCP. Instead,
our theory assumes that the normal state of the metal persists until
the critical point is reached.

%%%%%%%%

Our results demonstrate that at low temperatures the interference
between formally subleading composite modes $\sim \vecphi\cdot
\vecphi$ and disorder is ultimately more important than the direct
scattering off spin excitations $\vecphi$. Physically, this follows
from the singular behavior of the composite modes for small
momenta. Consequently, they affect the entire Fermi surface, in
contrast to the first-order scattering processes that are important
only around hot spots. Can this effect be observable in a specific
system? This depends on whether the other, competing phases, such as
superconductivity, intervene before the effects discussed here begin
to dominate the low-temperature transport.
%%%%%CHANGES
We believe that the effect of composite modes may be observable in the
electron-doped cuprates.
%%%%%
However, for a generic QCP the localizing
correction at lowest temperatures, followed by the positive
$\ln^2T$ correction in the intermediate temperature regime, are the
dominant contributions to the conductivity of the system.

\section*{Acknowledgments}
We are grateful to S.~V.~Syzranov for discussions during the early stages of this investigation
and to A.~D.~Mirlin
for a crucial comment
on the screening of the composite mode.
B.N.N. acknowledges support from the EU Network
Grant InterNoM.
%
%We also thank Alexander Mirlin
%for his crucial comment
%on the screening of the composite modes.
%
J.S. and P.W. acknowledge financial
support by the~Deutsche~Forschungsgemeinschaft through
Grant No.~SCHM~1031/4-1.

\appendix*

\section{Asymptotic behavior of the interaction correction}
	In this appendix we provide details of the evaluation of the conductivity formula in the asymptotic limit ${\cutoffe /T \gg 1}$.
	
	We approximate ${\tf^\prime(\cutoffe /(T \xz))/\xz \simeq \Theta_{\cutoffe /T-\xz}/\xz}$ as ${\tf^\prime(\cutoffe /(T \xz))/\xz}$ decays ${\sim 1/\xz^2}$ for ${\xz>\cutoffe/T}$ and ${\tf^\prime(\cutoffe /(T \xz))/\xz \approx 1}$ for ${\xz<\cutoffe/T}$.
	The frequency integral is approximated by
	\begin{equation} \label{eq:rescale-xx} 
	\begin{array}{rcccl}
%	\myintl_{0}^{\frac{\cutoffe}{T}} \dd \xx \dfrac{\tf^\prime(\xx)}{\xx} \D^A\left(\frac{\cutoffe}{T}\frac{1}{\xx}\right) 
%			& = & 
			\myintl_{1}^{\infty} \dd \xz \dfrac{\tf^\prime\left(\frac{\cutoffe}{T \xz} \right)}{\xz} \Ceff^A(\xz) 
			& \simeq & \myintl_{1}^{\cutoffe /T}  \dfrac{\dd\xz}{\xz} \Ceff^A(\xz) \,,
	\end{array} 
	\end{equation}
	
	(i) For ${T< T^\ast \ll \cutoffe}$ or ${\lambda \ln\left( \cutoffe /T \right) \gg 1}$,
	integration by parts over $\xz$ yields
	\begin{equation} \label{eq:zeta-int-by-parts} 
	\begin{array}{rl}
			& \myintl_{1}^{\cutoffe /T} \dfrac{\dd \xz}{\xz}\, 
						\frac{\left[ \lambda \left(\ln\xz - \frac{\I\pi}{2}\right) + F^{\rho}_0\right]}{1 -\left[\lambda\left(\ln\xz - \frac{\I\pi}{2}\right) + F^{\rho}_0\right] \tilde{\Pi}^A(\zz) } \\ \\
	\simeq 	& 
						\ln\left( \dfrac{\cutoffe}{T}\right) \left( -\dfrac{1}{\tilde{\Pi}^A(\zz)}\right) \\
			&			- \myintl_{1}^{\cutoffe /T} \dfrac{\dd \xz}{\xz}
						\frac{ -\left[ \lambda \left(\ln\xz - \frac{\I\pi}{2}\right) \right]}{\left\{1 -\left[\lambda\left(\ln\xz - \frac{\I\pi}{2}\right) \right] \tilde{\Pi}^A(\zz) \right\}^2 }
	\end{array} 
	\end{equation}
	In the asymptotic limit of ${\cutoffe/T \gg 1}$, the integral in the second line of \eqref{eq:zeta-int-by-parts} is proportional to $ \ln\left[ \ln (\cutoffe /T)\right]$.
	Thus, for low temperatures the dependence is governed by the $\ln T$ term.

	(ii) For ${T^\ast < T \ll \cutoffe}$ or ${\lambda \ln\xz < \lambda\ln\left( \cutoffe /T \right) \ll 1}$ we approximate the denominator by unity under the frequency integral
	\begin{widetext}
	\begin{equation} \label{eq:zeta-bare-prop} 
	\begin{array}{rcl}
	\myintl_{1}^{\cutoffe /T} \dfrac{\dd \xz}{\xz}\, 
			\dfrac{ 
			\left[ \lambda \left(\ln\xz - \frac{\I\pi}{2}\right) + F^{\rho}_0\right]
			}{
			1 -\left[\lambda\left(\ln\xz - \frac{\I\pi}{2}\right) + F^{\rho}_0\right] \tilde{\Pi}^A(\zz) 
			} 
			& \approx & 
						\myintl_{1}^{\cutoffe /T} \dfrac{\dd \xz}{\xz}\, 
						\left[ \lambda \left(\ln\xz - \frac{\I\pi}{2}\right) + F^{\rho}_0\right] 
						\\ \\
			& \simeq  &  \dfrac{\lambda}{2}\ln^2 \left( \dfrac{\cutoffe}{T}\right)
						+ \left(- \dfrac{\I\pi\lambda}{2} + F^{\rho}_0 \right) \ln \left(\dfrac{\cutoffe}{T} \right)
						
	\end{array} 
	\end{equation}
	\end{widetext}
	The logarithmic behavior of the bare propagator translates to the squared logarithmic temperature dependence
	while the constant contributions produce a subleading logarithmic temperature dependence.

%\bibliographystyle{apsrev} % to be replace 
%\bibliography{PRB-Interference-refs} % by bbl-file 

\end{document}